\documentclass[11pt,a4paper]{article}
\pdfoutput=1
\usepackage{jcappub}
\usepackage{mathtools}
\usepackage{amsfonts}
\usepackage{enumerate}
\usepackage{ragged2e}
\usepackage{afterpage}

\title{Discretising the velocity distribution for directional dark matter experiments}

\author[a,b]{Bradley J. Kavanagh}

\affiliation[a]{Institut de Physique Th\'{e}orique, CNRS, URA 2306 \& CEA/Saclay, F-91191 Gif-sur-Yvette, France}
\affiliation[b]{School of Physics \& Astronomy, University of Nottingham, University Park, Nottingham, NG7 2RD, UK}

\emailAdd{bradley.kavanagh@cea.fr}

\abstract{Dark matter (DM) direct detection experiments which are directionally-sensitive may be the only method of probing the full velocity distribution function (VDF) of the Galactic DM halo. We present an angular basis for the DM VDF which can be used to parametrise the distribution in order to mitigate astrophysical uncertainties in future directional experiments and extract information about the DM halo. This basis consists of discretising the VDF in a series of angular bins, with the VDF being only a function of the DM speed $v$ within each bin. In contrast to other methods, such as spherical harmonic expansions, the use of this basis allows us to guarantee that the resulting VDF is everywhere positive and therefore physical. We present a recipe for calculating the event rates corresponding to the discrete VDF for an arbitrary number of angular bins $N$ and investigate the discretisation error which is introduced in this way. For smooth, Standard Halo Model-like distribution functions, only $N=3$ angular bins are required to achieve an accuracy of around $10-30\%$ in the number of events in each bin. Shortly after confirmation of the DM origin of the signal with around 50 events, this accuracy should be sufficient to allow the discretised velocity distribution to be employed reliably. For more extreme VDFs (such as streams), the discretisation error is typically much larger, but can be improved with increasing $N$. This method paves the way towards an astrophysics-independent analysis framework for the directional detection of dark matter. }

\keywords{dark matter experiments, dark matter theory
\vskip13pt plus8pt minus11pt
\noindent{\bfseries\large\sffamily{Preprints:}} SACLAY-t15/016
}

\newcommand{\vmin}{v_\mathrm{min}}
\newcommand{\qhat}{\hat{\mathbf{q}}}

\newcommand{\acos}{\cos^{-1}}
\newcommand{\asin}{\sin^{-1}}
\newcommand{\atan}{\tan^{-1}}
\newcommand{\kms}{\textrm{ km s}^{-1}}
\newcommand{\dbd}[2]{\ifmmode \frac{\textrm{d}#1}{\textrm{d}#2}\else $\textrm{d}#1/\textrm{d}#2$\fi}

\begin{document}
\begin{center}
\hfill
\footnotesize{SACLAY-t15/016}
\end{center}
    \maketitle

\section{Introduction}

The measurement of the directionality of nuclear recoils in dedicated low-background detectors has long been proposed as a method to detect particle dark matter (DM) \cite{Spergel:1988}. The motion of the Sun through the Galactic DM halo generates a so-called `wind' of DM particles, which would appear to originate from the direction of the constellation of Cygnus. The resulting recoil spectrum will be peaked in the opposing direction (or for high mass DM, in a ring around this direction \cite{Bozorgnia:2011}). Such a directional signal would be a smoking gun for DM against the typically isotropic distribution of residual backgrounds \cite{Copi:1999, Copi:2002,Morgan:2005,Billard:2010c,Green:2010b}. Directional detection may also provide a way to circumvent the neutrino floor \cite{Grothaus:2014} and to probe the local structure of the DM velocity distribution \cite{Billard:2013, Ohare:2014}.

A number of directional experiments are currently in the prototype stage. The predominant technology is the low-pressure time-projection chamber (TPC), in which nuclear recoils leave an $\mathcal{O}$(1 mm) track of ionised electrons, which can be imaged by drifting the electrons (or electron-transport gas) to an anode grid. This technology is currently utilised by the DRIFT \cite{Daw:2012,Battat:2014}, MIMAC \cite{Riffard:2013,Santos:2013}, DMTPC \cite{Monroe:2012, Battat:2013}, NEWAGE \cite{Miuchi:2010, Miuchi:2012} and D3 \cite{Vahsen:2012} collaborations and the problem of reconstructing the recoil direction using this technology has been much studied \cite{Burgos:2007, Billard:2012}. More recent proposals for directional detectors include nuclear emulsions \cite{Naka:2012}, electronic scattering in crystals \cite{Essig:2012b},  DNA-based methods \cite{Drukier:2012} and columnar recombination in Xenon targets \cite{Nygren:2013nda, Mohlabeng:2015efa}.

Directional experiments such as these may be the only possibility for probing the full 3-dimensional velocity distribution of DM, $f(\mathbf{v})$, which is \textit{a priori} unknown. The standard analysis of direct detection experiments assumes the so-called Standard Halo Model (SHM), a simplified model for the DM halo as an isotropic, isothermal sphere of particles \cite{Green:2012}. However, such a simple description is unlikely to be an accurate description of the Milky Way halo \cite{Evans:2000,Widrow:2000,Lisanti:2010,Bhattacharjee:2012,Fornasa:2013}. In addition, $N$-body simulations raise the possibility of more complicated distributions, including debris flows \cite{Lisanti:2013, Kuhlen:2012}, tidal streams \cite{Freese:2004, Freese:2005} or a dark disk \cite{Read:2009, Read:2010,Pillepich:2014}. The impact of these astrophysical uncertainties on directional signals has previously been studied \cite{Bozorgnia:2011, Billard:2013, Ohare:2014}. While non-standard astrophysics should not severely affect the `smoking gun' directional signature of DM recoils \cite{Billard:2013}, it may still pose a potential problem for future data. In particular, trying to reconstruct the DM parameters (such as mass $m_\chi$ and interaction cross section $\sigma$) from a potential signal requires some assumptions to be made about the form of $f(\mathbf{v})$. As with non-directional detection techniques, poor assumptions about the astrophysics of DM can lead to biased reconstructions of these parameters \cite{Peter:2011,Fairbairn:2012, Kavanagh:2012, Kavanagh:2013a, Kavanagh:2014}. 

Astrophysics-independent approaches to \textit{non-directional} experiments have received significant attention in the past few years, including so-called `halo-independent' methods \cite{Fox:2011b, Fox:2011c, Frandsen:2012, Gondolo:2012, DelNobile:2013a, Bozorgnia:2013, DelNobile:2013b, Feldstein:2014a, Feldstein:2014b, Fox:2014, Cherry:2014} and attempts at parametrising the 1-dimensional speed distribution \cite{Strigari:2009, Peter:2011, Kavanagh:2012, Kavanagh:2013a}. There have also been several attempts to construct a suitable parametrisation for the 3-dimensional velocity distribution, relevant for directional experiments \cite{Billard:2010b, Lee:2012, Alves:2012,Lee:2014}. This is significantly more difficult than in the non-directional case, owing to the presence of the angular components of $f(\mathbf{v})$. A single 1-dimensional function is therefore no longer sufficient to describe the full distribution function and a suitable angular basis must be found. Several bases have been suggested (e.g.~Refs.~\cite{Alves:2012, Lee:2014}) but all typically suffer from the same problem, which has not previously been addressed in the literature: they allow $f(\mathbf{v})$ to take negative - and therefore unphysical - values, which can lead to spurious results. We discuss this problem in more detail in Sec.~\ref{sec:problems}.

In this work, we present a framework which allows the velocity distribution to be parametrised in a model-independent way, while guaranteeing that it is everywhere positive. This consists of an angular discretisation of the velocity distribution, decomposing it into a series of 1-dimensional functions, each of which is constant over a given bin in the angular coordinates. This is motivated in the first instance by its ability to describe the simplest directional signal, a forward-backward asymmetry in the scattering rate. However, we extend the discretisation to an arbitrary number of bins $N$ in the angular variables, such that in the limit of large $N$, the full 3-dimensional distribution can be recovered. For arbitrary $N$, we also demonstrate how to calculate the Radon transform (the function which encodes the angular dependence of the scattering rate) and therefore how to compute the full scattering rate.

Here, we focus on the angular discretisation itself, and do not attempt a full reconstruction of the DM velocity distribution and particle physics parameters (as in e.g.~\cite{Peter:2011, Kavanagh:2012}). Instead, we investigate the magnitude of the error introduced by this angular discretisation for two different benchmark velocity distributions. To do this, we compare the event rate within each angular bin obtained using the full and discrete velocity distributions. Qualitatively, we discuss and compare the energy dependence of the event rate within each bin. Quantitatively, we can calculate the number of signal events in each angular bin for the full and discrete distributions, and compare the result. This allows us to evaluate whether (and in which scenarios) fitting to data using such an angular discretisation gives accurate results.

In order to perform this analysis, we assume that the velocity distribution is known exactly. In the analysis of real experiments, a parametrisation of $f(\mathbf{v})$ within each bin must be chosen and fit to the data. We leave the choice of a suitable parametrisation of the remaining 1-dimensional functions of $v$ to future work, meaning that the results presented here represent a lower limit on the error induced using the discretised velocity distribution. However, we lay out a framework which can be used in the analysis of future data to extract coarse-grained information about the DM velocity distribution in a way which was not previously possible.

In Sec.~\ref{sec:formalism}, we present the directional detection formalism, including calculation of the scattering cross section and the Radon transform, and a brief discussion of the associated astrophysical uncertainties. This is followed in Sec.~\ref{sec:problems} by a discussion of previous attempts to overcome these uncertainties. In Sec.~\ref{sec:discretisation}, we present the discretised velocity distribution which is the focus of this work. In Sec.~\ref{sec:compare}, we investigate how this discretised approximation compares to the exact result, both at the level of the Radon transform and the event rate. Finally, we discuss the significance of these results and plans for future work in Sec.~\ref{sec:discussion}. The procedure for calculating the Radon transform from the discretised distribution is presented in Appendix~\ref{app:Radon}, while we relegate the full derivation of this formula to Appendix~\ref{app:RadonDeriv}.

\section{Directional formalism}
\label{sec:formalism}
The directional rate per unit detector mass for recoils of energy $E_R$ can be written as \cite{Gondolo:2002}

\begin{equation}
\label{eq:Rate}
\frac{\mathrm{d}R}{\mathrm{d}E_R\mathrm{d}\Omega_q} = \frac{\rho_0}{4\pi \mu_{\chi p}^2 m_\chi} \sigma^p \mathcal{C}_N F^2(E_R) \hat{f}(v_\textrm{min}, \hat{\mathbf{q}}) \,,
\end{equation}
where $\hat{\mathbf{q}} = (\sin\theta\cos\phi, \sin\theta\sin\phi, \cos\phi)$ is a unit vector pointing in the direction of the recoil. 
We have written the rate $R$ in terms of the DM-proton cross section at zero-momentum transfer $\sigma_p$, which may be spin-independent or spin-dependent; the form factor $F^2(E_R)$, which describes the loss of coherence due to the finite size of the nucleus; and an enhancement factor $\mathcal{C}_N$, which describes the enhancement of the rate for a nucleus $N$ relative to the DM-proton rate. The local DM density is written as $\rho_0$, the DM mass as $m_\chi$ and the reduced DM-proton mass as $\mu_{\chi p} = m_\chi m_p/(m_\chi + m_p)$.

For the gaseous targets used in low-pressure TPCs, relatively light nuclei are typically used. For example, the DRIFT experiment \cite{Daw:2012,Battat:2014} uses $\mathrm{CF}_4$ as the target gas, with spin-dependent (SD) scattering from $^{19}\mathrm{F}$ nuclei expected to be the dominant interaction with DM particles. For SD interactions, the enhancement factor can be written as \cite{Cerdeno:2010,Cannoni:2013}

\begin{equation}
\mathcal{C}_N^{SD} = \frac{4}{3}\frac{J+1}{J} \left|\langle S_p \rangle + a_n/a_p \langle S_n \rangle\right|^2\,,
\end{equation}
where $J$ is the total nuclear spin, $\langle S_{p,n} \rangle$ is the expectation value of the proton and neutron spin in the nuclear ground state with maximal magnetic quantum number, and $a_n/a_p$ is the ratio of the DM-neutron and DM-proton couplings. The ratio $a_n/a_p$ depends on the specific model of DM under consideration, although simplifying assumptions are often used, including pure-nucleon couplings ($a_p = 0$ or $a_n = 0$) \cite{Tovey:2000} or values motivated by specific models (e.g. $a_n/a_p = \pm 1$) \cite{Vasquez:2012}. 

The SD form factor can be written in terms of the spin structure functions of the nucleus $S_{ij}(q)$, which depend on the momentum transfer $q = \sqrt{2 m_N E_R}$, for a nucleus of mass $m_N$:

\begin{equation}
F^2_{SD}(q) = S(q)/S(0)\,,
\end{equation}
with
\begin{equation}
S(q) = a_0^2S_{00}(q) + a_0 a_1 S_{01}(q) + a_1^2 S_{11}(q)\,.
\end{equation}
The isoscalar and isovector couplings are related to the proton and neutron couplings by $a_0 = a_p + a_n$ and $a_1 = a_p - a_n$ \cite{Cannoni:2013}. The spin structure functions (and the proton and neutron spin matrix elements) can be calculated using nuclear shell models (see e.g.~Refs.~\cite{Ellis:1988,Engel:1989,Iachello:1991,Ressel:1993, Dimitrov:1995, Menendez:2012}). Here, we focus on Fluorine targets, assuming $a_p = a_n$, with spin structure functions and matrix elements taken from Ref.~\cite{Divari:2013}. The spin structure functions lead to a roughly exponential suppression of the recoil spectrum as a function of recoil energy.

The DM velocity distribution enters into the scattering rate through the Radon transform
\begin{equation}
\hat{f}(\vmin, \qhat) = \int_{\mathbb{R}^3}  f(\mathbf{v}) \delta \left(\mathbf{v} \cdot \qhat - \vmin\right) \mathrm{d}^3 \mathbf{v}\,,
\end{equation}
where $\vmin$ is the minimum DM speed required to excite a nuclear recoil of energy $E_R$:
\begin{equation}
\label{eq:vmin}
v_\mathrm{min} = v_\mathrm{min}(E_R) = \sqrt{\frac{m_N E_R}{2 \mu_{\chi N}^2}}\\.
\end{equation}
Geometrically, the Radon transform corresponds to an integral over $f(\textbf{v})$ on a plane perpendicular to $\hat{\textbf{q}}$, which has a perpendicular distance $v_\mathrm{min}$ from the origin. Physically, this corresponds to integrating over all DM velocities which satisfy the kinematic constraints for exciting a nuclear recoil of energy $E_R$ in direction $\hat{\mathbf{q}}$.

The typical assumption for the form of the DM velocity distribution is the Standard Halo Model (SHM). For a spherically symmetric, isothermal DM halo, with density profile $\rho(r) \propto r^{-2}$, the resulting velocity distribution has a Maxwell-Boltzmann form in the reference-frame of the Galaxy, \cite{Gondolo:2002}
\begin{equation}
f(\mathbf{v}) = \frac{1}{(2\pi \sigma_v^2)^{3/2}} \exp\left(-\frac{\mathbf{v}^2}{2\sigma_v^2}\right)\,,
\end{equation}
with velocity dispersion $\sigma_v$. The corresponding Radon transform also takes the form of a Gaussian,
\begin{equation}
\label{eq:analRadon}
\hat{f}(\vmin, \hat{\mathbf{q}}) = \frac{1}{(2\pi \sigma_v^2)^{1/2}} \exp\left(-\frac{\vmin^2}{2\sigma_v^2}\right)\,.
\end{equation}
Using the properties of the Radon transform we can obtain the corresponding result in the Earth frame by performing a Galilean transformation  $\mathbf{v} \rightarrow \mathbf{v} + \mathbf{v}_e$, with $\mathbf{v}_\mathrm{e}$ the velocity of the Earth with respect to the rest frame of the halo, in which case

\begin{equation}
\hat{f}(\vmin, \hat{\mathbf{q}}) \rightarrow \hat{f}(\vmin + \mathbf{v}_\mathrm{e}\cdot \hat{\mathbf{q}}, \hat{\mathbf{q}})\,.
\end{equation}
The SHM velocity distribution in the laboratory frame is then
\begin{equation}
\label{eq:SHM_Radon}
\hat{f}(\vmin, \hat{\mathbf{q}}) = \frac{1}{(2\pi \sigma_v^2)^{1/2}} \exp\left(-\frac{(\vmin +  \mathbf{v}_\mathrm{e}\cdot\hat{\mathbf{q}}) ^2}{2\sigma_v^2}\right)\,.
\end{equation}
Within the SHM, one typically assumes values of $v_\mathrm{e} \approx 220 \kms$ and $\sigma_v = v_\mathrm{e}/\sqrt{2} \approx 156 \kms$. However, there are still uncertainties on these values of at least $10\%$ \cite{Feast:1997, Schonrich:2012, Bovy:2012a}. Introducing a cut off in the distribution at the Galactic escape speed $v_\mathrm{esc}$ introduces further uncertainty into the model \cite{Rave:2014}.

Despite its widespread use, the SHM is unlikely to be an accurate representation of the DM halo. Observations and N-body simulations indicate that the halo should deviate from a $1/r^2$ profile and may not be spherically symmetric. As a result alternative models have been proposed. Speed distributions associated with triaxial halos \cite{Evans:2000} or with more realistic density profiles \cite{Widrow:2000} have been suggested, as well as analytic parametrisations which should provide more realistic behaviour at low and high speeds \cite{Lisanti:2010}. Self-consistent distribution functions reconstructed from the potential of the Milky Way have also been obtained \cite{Bhattacharjee:2012,Fornasa:2013}.

It is also possible to extract the speed distribution from N-body simulations. Such distribution functions tend to peak at lower speeds than the SHM and have a more populated high speed tail \cite{Vogelsberger:2009, Kuhlen:2010, Mao:2012}. There are also indications that DM substructure may be significant, causing `bumps' in the speed distribution, or that DM which has not completely phase-mixed - so-called `debris flows' - may have a contribution \cite{Kuhlen:2012}.

Another result obtained from simulations is the possibility of a dark disk. When baryons are included in simulations of galaxy formation, this can result in DM subhalos being preferentially dragged into the disk plane where they are tidally stripped \cite{Read:2009, Read:2010}. The result is a dark disk which corotates with the stellar disk, though with a smaller velocity dispersion, typically $\sigma_v^{DD} \sim 50 \kms$. Such a dark disk could comprise a large fraction (up to 50\%) of the total DM density, though more recent results \cite{Pillepich:2014} suggest a smaller dark disk contribution (around 10\%), depending on the merger history of the Milky Way.

Finally, direct detection experiments probe the DM halo on sub-milliparsec scales and there is therefore the possibility that DM sub-structures could dominate the local distribution. Analyses of N-body simulations suggest that no individual subhalos should dominate the local density \cite{Helmi:2002,Vogelsberger:2007}. However, local structures could still be significant, especially streams of DM from the tidal disruption of Milky Way satellites, such as Sagittarius \cite{Freese:2004, Freese:2005, Savage:2006}. 

Such a wide range of possibilities for the form of $f(\mathbf{v})$ leads to substantial uncertainties on the predicted event rate for directional detectors. This in turn can lead to potential bias in the reconstruction of other DM parameters, if these uncertainties are not properly accounted for.

\section{Problems with parametrising the velocity distribution}
\label{sec:problems}

With promising developments in directional detector technology, it is interesting to ask what information about the velocity distribution we could, in principle, extract from a directional signal. Early attempts to address this question involved extending the SHM to allow anisotropic velocity dispersions and fitting these new parameters to mock data \cite{Billard:2010b}. The possibility of including additional structures, such as streams and dark disks, has also been considered \cite{Lee:2012}. However, such methods are likely to be accurate only if the underlying velocity distribution is well-described by the chosen model.

Alves, Hendri and Wacker \cite{Alves:2012} investigated the more general possibility of describing $f(\textbf{v})$ in terms of a series of special functions of integrals of motion (energy and angular momentum). These can then be fit to data, with around 1000 events required to distinguish between the SHM and a Via Lactea II distribution function \cite{Kuhlen:2008}. However, the special, separable form of the velocity distribution requires that the dark matter halo is in equilibrium. More troubling is that the resulting velocity distribution is not guaranteed to be everywhere positive and therefore not all combinations of parameters correspond to physical distribution functions.

A more general parametrisation for the velocity distribution was recently proposed by Lee \cite{Lee:2014}. In this approach, the velocity distribution is decomposed into products of Fourier-Bessel functions and spherical harmonics. This is completely general and does not require assumptions about the halo being in equilibrium. Lee also gives an analytic expression for the Radon transform of the Fourier-Bessel basis, making this approach computationally efficient. However, this approach can also produce negative-valued, non-physical distribution functions, as in the case of the method of Alves et al..

In fact, any decomposition in terms of spherical harmonics leads to this problem, because the spherical harmonic basis functions can have negative values. It is unclear how this issue will affect parameter reconstruction. It may lead to parameter estimates (e.g.~for $m_\chi$ or $\sigma_p$) which are unphysical, as they require an unphysical distribution function to fit the data well. Without some criteria which determines which coefficients of the spherical harmonics lead to strictly positive distribution functions, it may not be possible to reject such parameter points. We may attempt to numerically test each parametrised distribution function for negative values but for a real function of three parameters $f(\textbf{v}) = f(v_x, v_y, v_z)$ this would require a very large number of evaluations, which may not be computationally feasible (and does not absolutely guarantee positive-definiteness). In addition, physical distributions may occupy only a small fraction of the total space of parameters making parameter sampling and reconstruction difficult. 

Even if positive-definiteness could be ensured, it is not clear how to interpret the parameters reconstructed in this way. Spherical harmonic approximations of typical velocity distributions such as the SHM (obtained by integrating out the coefficients of the basis functions) tend to produce distributions which do contain negative values. This is especially true when only a small number of basis functions is used. However, the `true' spherical harmonic coefficients obtained in this way cannot be obtained by fitting to data (because we would reject those which lead to negative values). This may again lead to a bias in the reconstructed DM parameters, because the `true' values do not lie in the allowed parameter space. Such potential problems have not previously been noted in the literature.

In order to fit to data, then, it is necessary to decompose $f(\textbf{v})$ into a series of angular components $A^i$:
\begin{equation}
f(\textbf{v}) =  f(v, \cos\theta', \phi')=  f^1(v) A^1(\theta',\phi') + f^2(v) A^2(\theta',\phi') +f^3(v) A^3(\theta',\phi') +...\,.
\end{equation}
We then truncate the series at some order, leaving only a finite number of 1-dimensional functions $f^i(v)$ which are unknown. This reduces the problem of attempting to fit a function of the 3-dimensional variable $\textbf{v}$ to the problem of parametrising a series of 1-dimensional functions, which is much more tractable. Of course, we should be careful that this truncation preserves enough angular information to still provide a good approximation to $f(\textbf{v})$. However, as more data becomes available, we can add more terms to the series to capture more angular features in the distribution.

As we have discussed, the spherical harmonic basis may not be an appropriate choice for this decomposition. In the next section, I will present an alternative decomposition which can guarantee that the velocity distribution is everywhere positive and therefore represents a promising and general method for extracting information from directional experiments.

\section{A discretised velocity distribution}
\label{sec:discretisation}

In order to ensure that the velocity distribution is everywhere positive, we propose that the velocity distribution be discretised into $N$ angular components:

\begin{equation}
\label{eq:discretisedf}
f(\textbf{v}) = f(v, \cos\theta', \phi') =
\begin{cases}
f^1(v) & \textrm{ for } \theta' \in \left[ 0, \pi/N\right]\,, \\
f^2(v) & \textrm{ for } \theta' \in \left[ \pi/N, 2\pi/N\right]\,, \\
 & \vdots\\
f^k(v) & \textrm{ for } \theta' \in \left[ (k-1)\pi/N, k\pi/N\right]\,, \\
 & \vdots\\
f^N(v) & \textrm{ for } \theta' \in \left[ (N-1)\pi/N, \pi\right]\,. \\
\end{cases}
\end{equation}
Over each bin in $\theta'$, $f(\textbf{v})$ has no angular dependence and depends only on a single function of the DM speed. The resulting velocity distribution will be everywhere positive, as long as a suitable parametrisation for the $f^k(v)$ is chosen which is itself everywhere positive.

We consider for simplicity only a discretisation in $\cos\theta'$. We note that in this work we will only be considering the azimuthally-averaged event rate (i.e.~integrated over the recoil angle $\phi$). In this case, we emphasise that no assumptions about the $\phi'$-dependence of $f(\mathbf{v})$ are required, as the integral over $\phi'$ can be performed exactly, and the azimuthally-averaged rate depends only on the azimuthally-averaged velocity distribution (see Appendix~\ref{app:RadonDeriv}). We can therefore take the velocity distribution to be independent of $\phi'$ without loss of generality.  However, this analysis could be extended to consider bins in the $\phi$ angle, with an additional discretisation in $\phi'$ if required. 

The motivation for this discretised description is that the simplest signal (beyond an isotropic $N=1$ signal) which can be observed with a directional detector is an asymmetry between the event rates in, say, the forward and backward directions. Shortly after the confirmation of a dark matter signal at a directional detector, the number of events may still be quite small (for example, the roughly 10 events required to distinguish from an isotropic background \cite{Morgan:2005}, or 30 events required to confirm the peak recoil direction \cite{Green:2010b}). In this small statistics scenario, constraining a large number of free functions is not feasible. However, if we discretise $f(\textbf{v})$ into $N=2$ angular components, it may be possible to extract some meaningful directional information with only a small number of events. With larger numbers of events, $N$ can be increased to allow more directional information to be extracted.

\begin{figure}[ht!]
  \centering
  \vspace{1.5cm}
  \includegraphics[width=0.49\textwidth]{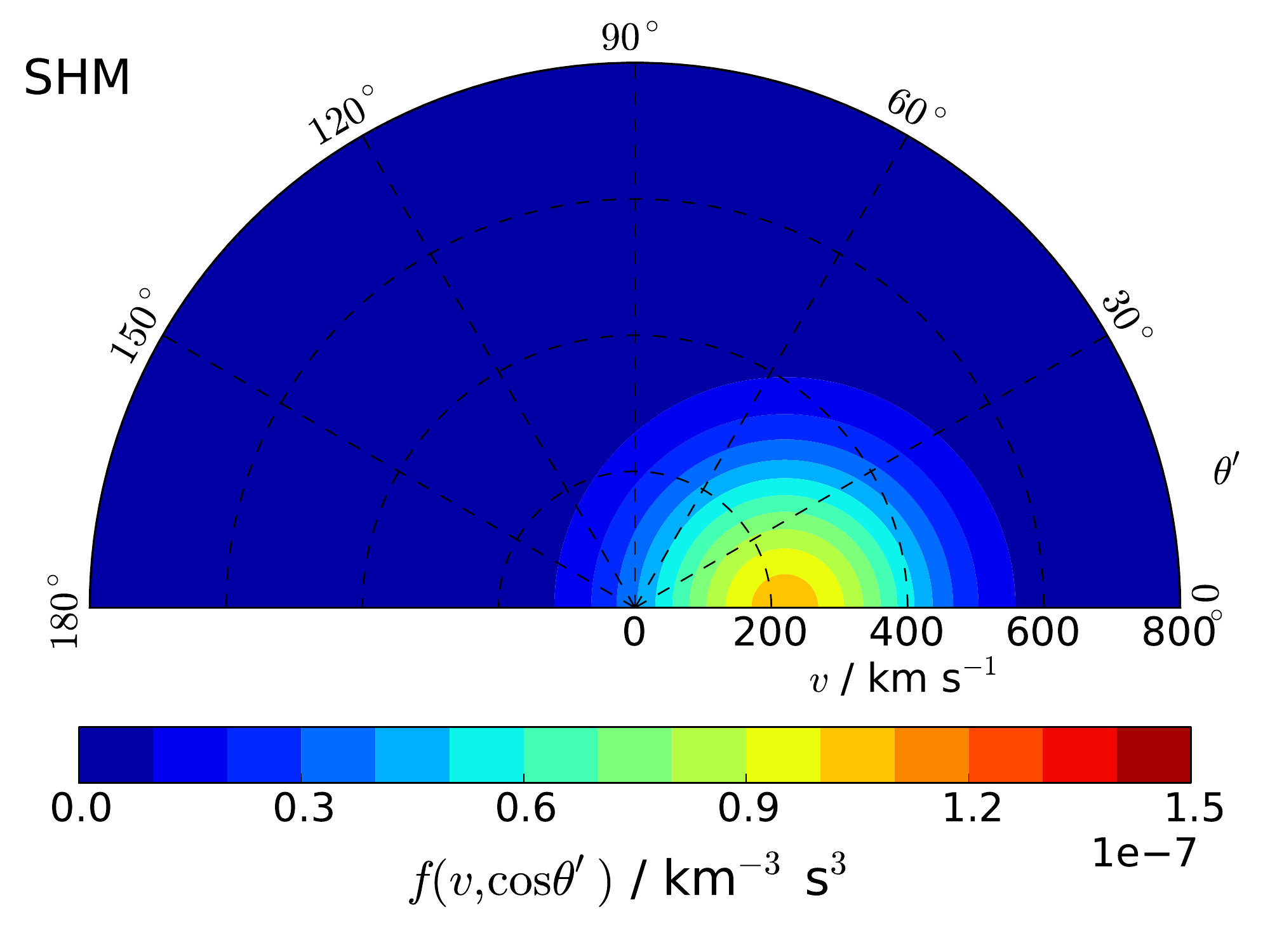}
  \includegraphics[width=0.49\textwidth]{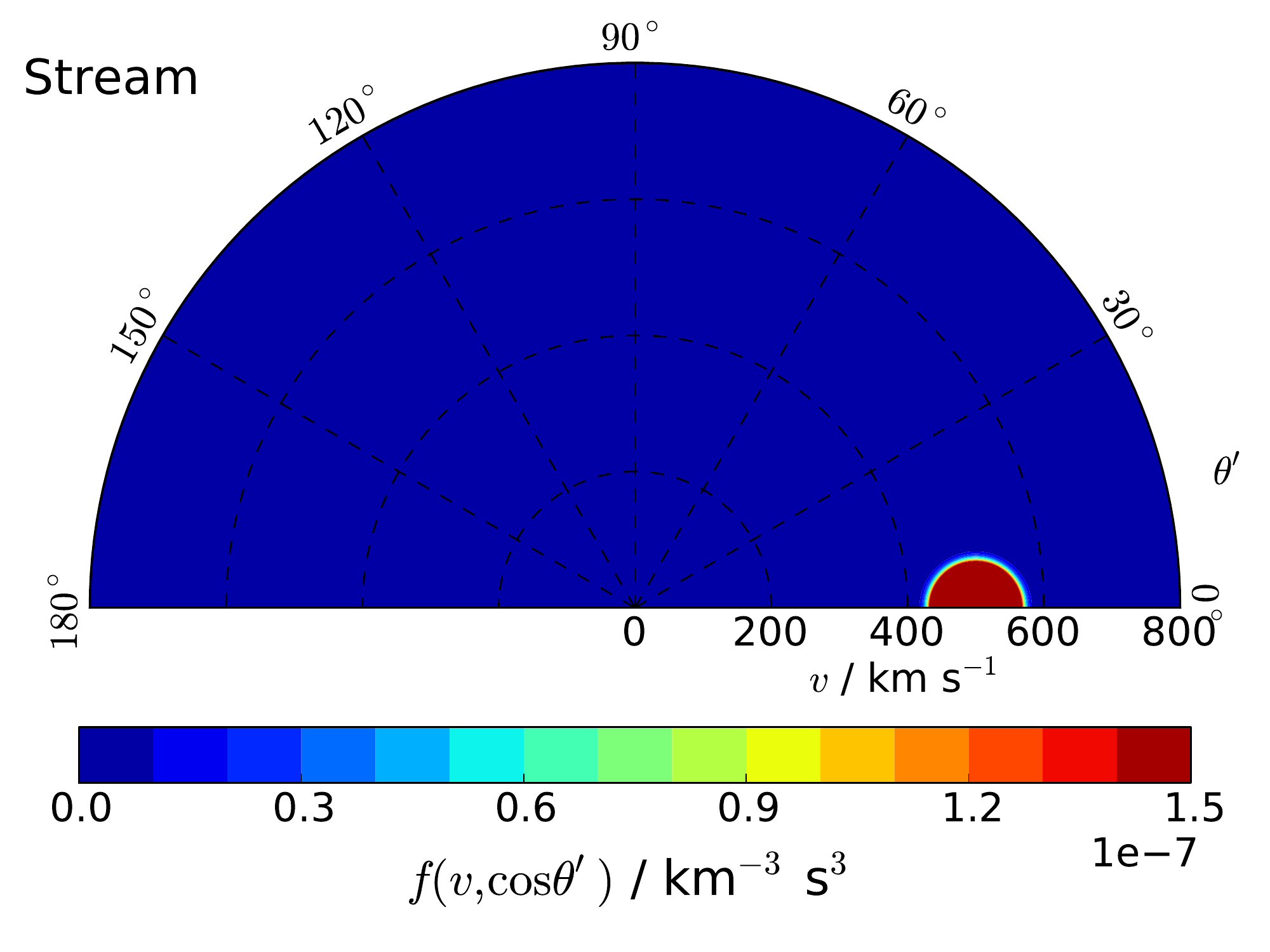}  
  
  \includegraphics[width=0.49\textwidth]{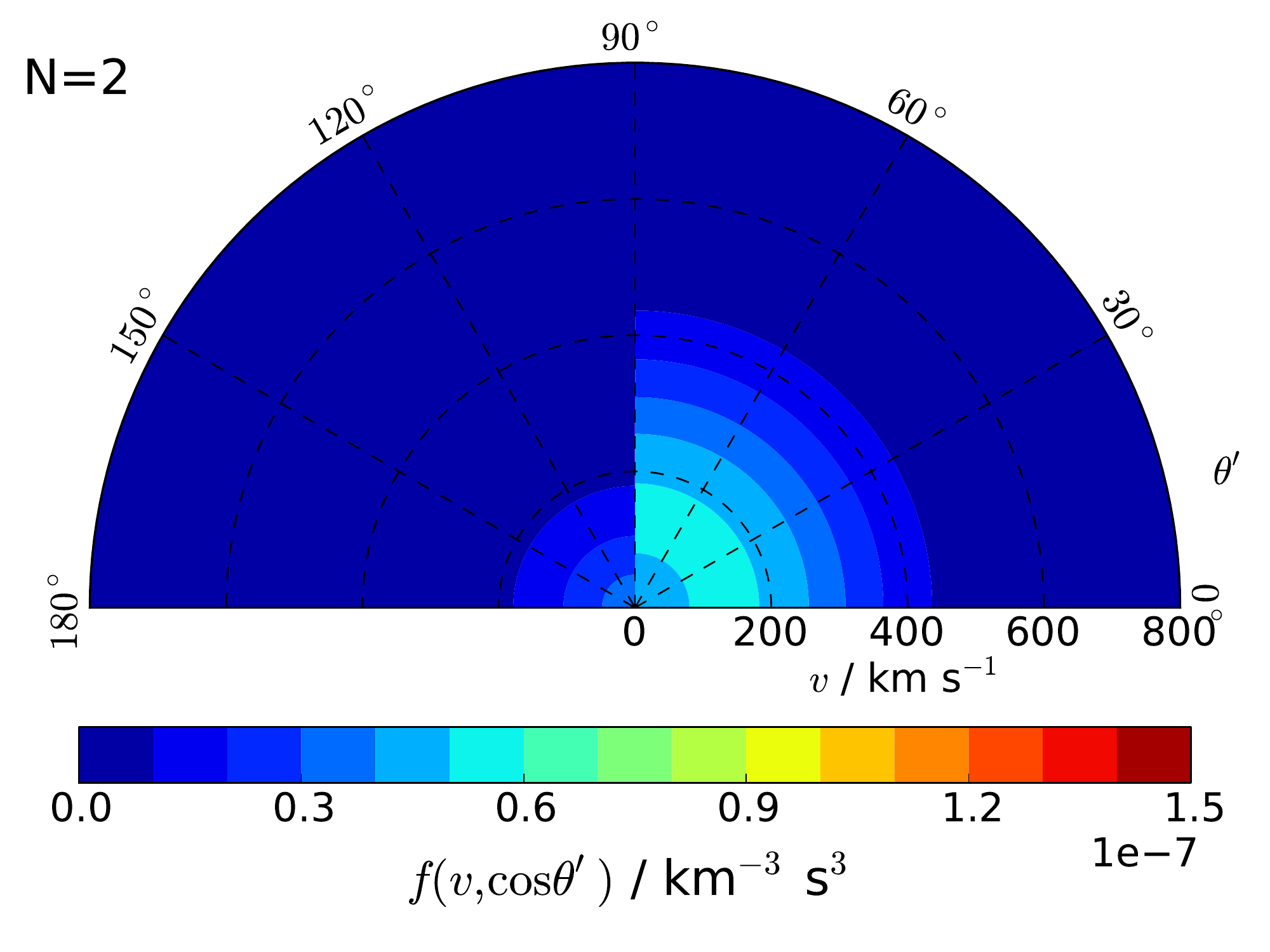}
  \includegraphics[width=0.49\textwidth]{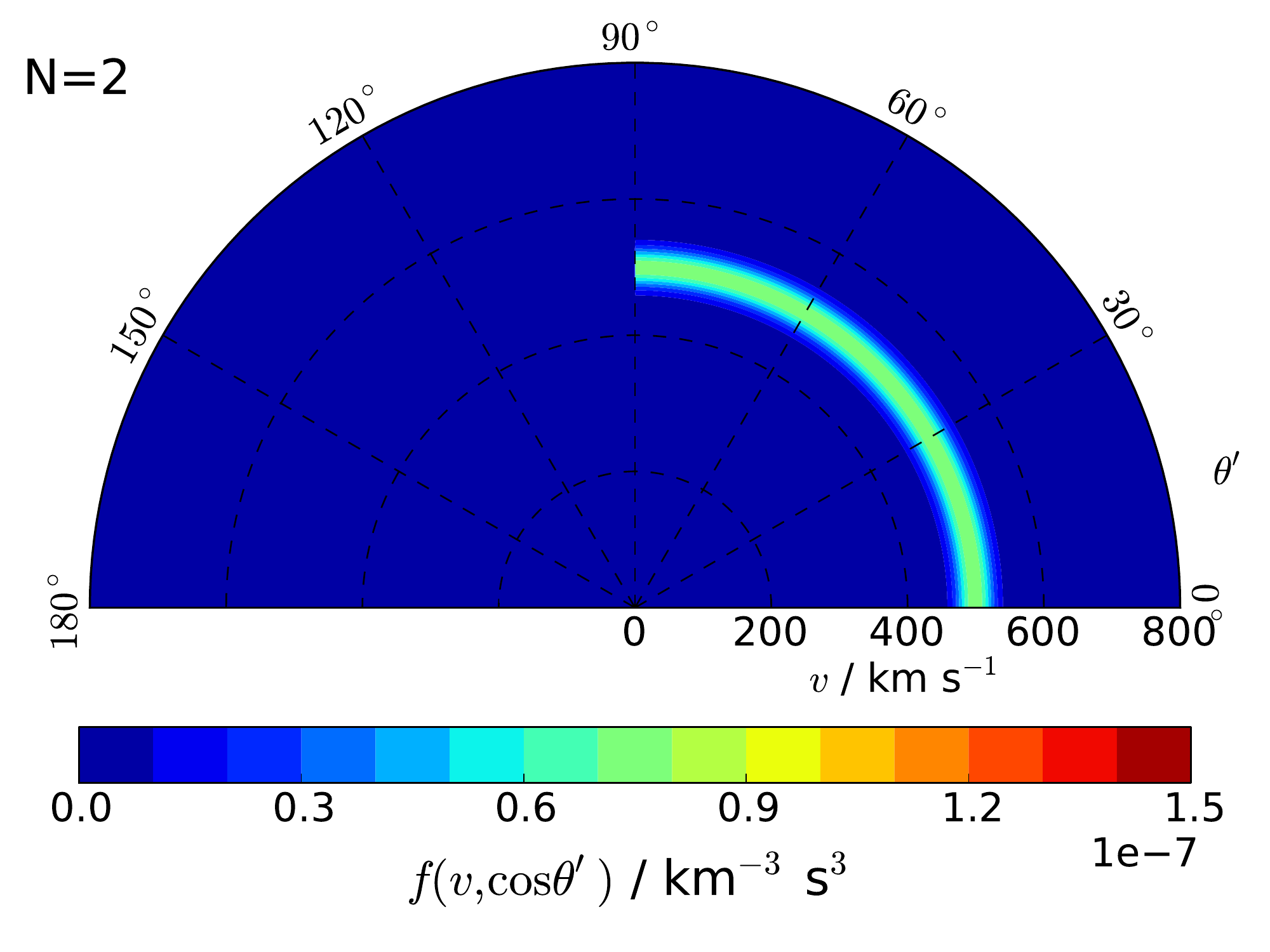}
   
  \includegraphics[width=0.49\textwidth]{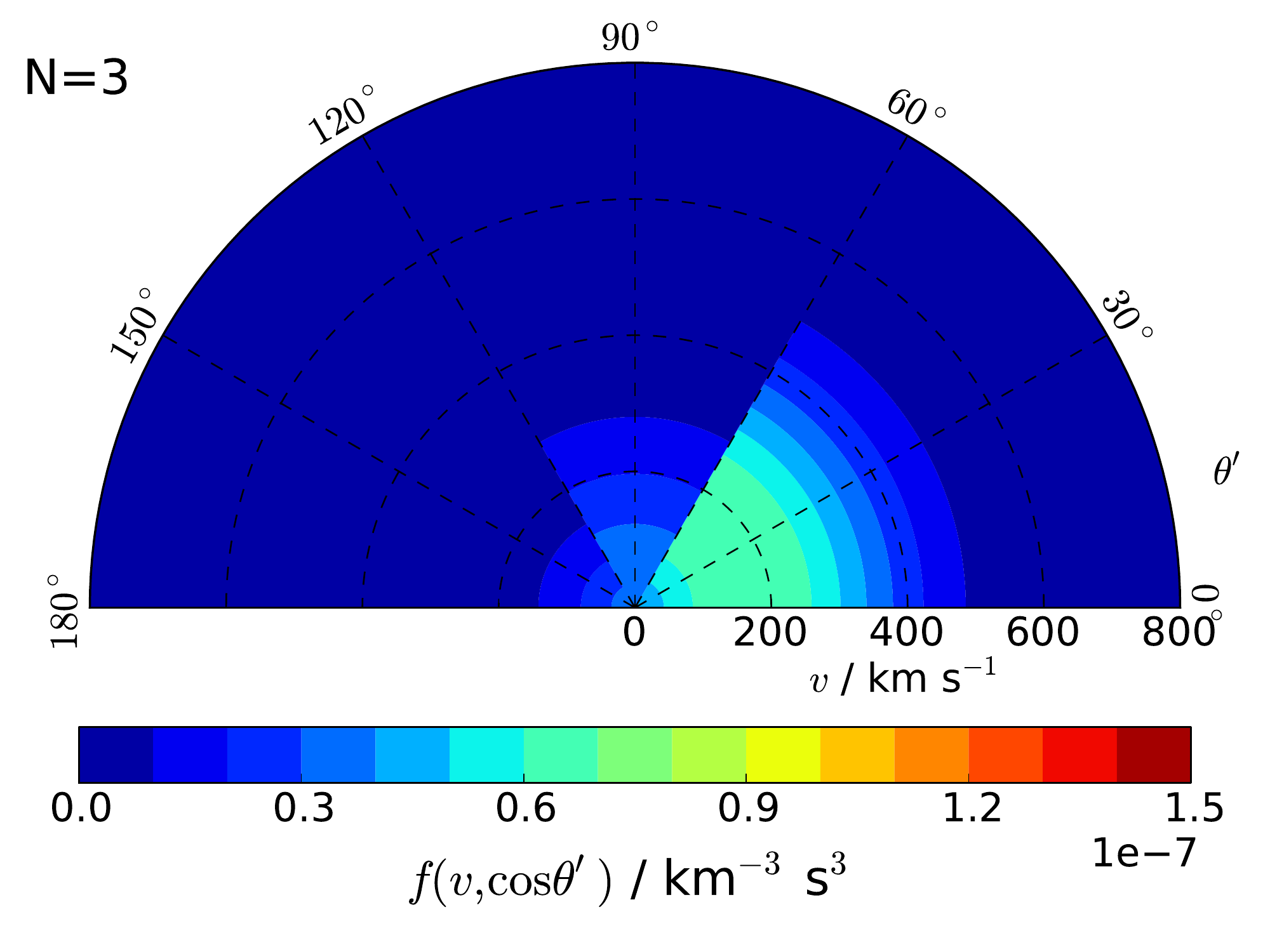}
  \includegraphics[width=0.49\textwidth]{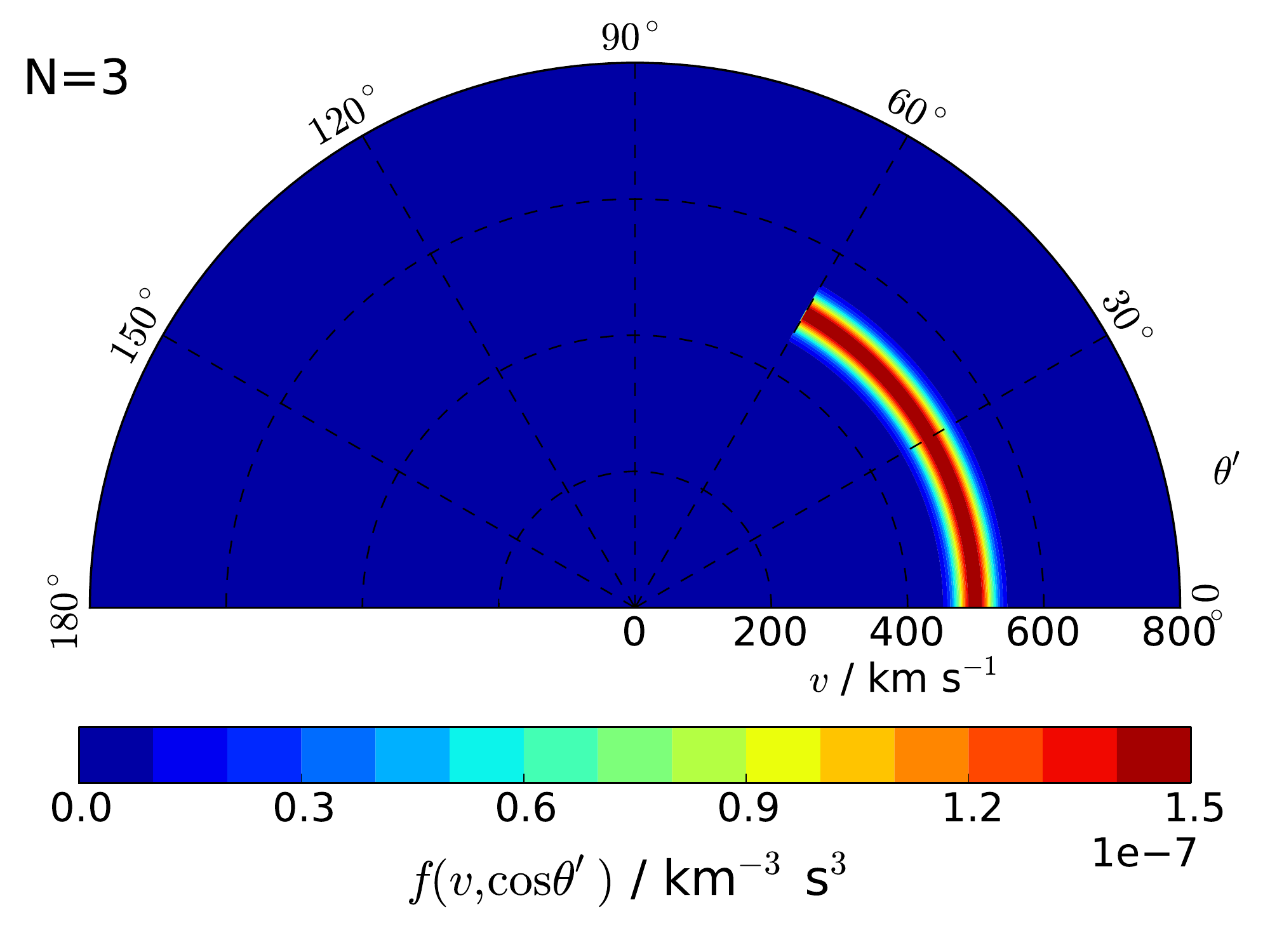}
\caption{The full velocity distribution (top) as well as $N=2$ (middle) and $N=3$ (bottom) discretised approximations for two examples: the SHM with $v_\mathrm{e} = 220 \kms$, $\sigma_v = 156 \kms$ (left column) and a stream with $v_\mathrm{e} = 500 \kms$, $\sigma_v = 20 \kms$ (right column). In each case, we have integrated over the $\phi'$ direction and only show $f(v, \cos\theta')$. The vector $\textbf{v}_\textrm{e}$ is aligned along $\theta' = 0$. The same colour scale is used in each plot. In the case of the full stream distribution (top right), large values have been truncated for easier comparison with the other plots. }
  \label{fig:Discrete}
  \afterpage{\clearpage}
\end{figure}

We show in Fig.~\ref{fig:Discrete} some examples of this discretised velocity distribution. We show the SHM velocity distribution (top left), as well as the $N=2$ (middle left) and $N=3$ (bottom left) discretised approximations, in all cases integrated over $\phi'$. These approximations are obtained by averaging the full velocity distribution over each bin in $\theta'$:

\begin{equation}
\label{eq:averagef}
f^{k}(v) = \frac{1}{\cos((k-1)\pi/N) - \cos(k\pi/N)}\int_{\cos(k\pi/N)}^{\cos((k-1)\pi/N)} f(\mathbf{v}) \, \mathrm{d}\cos\theta'\,.
\end{equation}
For comparison, the results for a stream distribution are also shown in the right column. We describe these two distribution functions in more detail at the start of Sec.~\ref{sec:compare}.

Upon discretisation, the distribution which is initially focused in one direction (towards $\theta' = 0^\circ$) now becomes constant over $\theta'$ in each bin. For the SHM (left column), there is a predominantly forward-going component ($\cos\theta' > 0$), as well as a smaller backwards component ($\cos\theta' < 0$). For the stream (right column), which has a much narrower dispersion and higher peak speed, the $N=2$ discretised distribution is almost entirely in the forward direction. However, we note that due to the discretisation, there is now a sizeable population of particles with transverse velocities ($\theta' = 90^\circ$), which was not the case in the full distribution. In the $N=3$ discretisation (bottom row), the velocity distributions become more focused and begin to recover more of the directionality of the full distributions. In the limit of large $N$, the full distribution can be recovered.

In order to determine the corresponding event spectrum, we must calculate the Radon transform from this discretised $f(\mathbf{v})$. We introduce the integrated Radon transform (IRT), integrated over the same angular bins as for $f(\mathbf{v})$:

\begin{equation}
\label{eq:discreteRadon}
\hat{f}^j(v_\textrm{min}) = \int_{\phi = 0}^{2\pi} \int_{\cos(j\pi/N)}^{\cos((j-1)\pi/N)} \hat{f}(v_\textrm{min}, \hat{\textbf{q}})\, \mathrm{d}\cos\theta\mathrm{d}\phi\,,
\end{equation}
where $\hat{\mathbf{q}} = (\sin\theta\cos\phi, \sin\theta\sin\phi, \cos\phi)$.\footnote{The integration over $\phi$ essentially reduces the experiment to what is referred to as 1-dimensional directional detection, in which only information about $\theta$ is available. However, it has been shown that this does not significantly reduce the discovery power of the detector when compared with 3-dimensional detection \cite{Billard:2015}.} In principle, we can define the IRT over a set of bins different from those used to define the discretised distribution function. However, using the same bins in both cases typically simplifies the calculation of the IRT. A notable exception would be to consider fewer bins for the Radon transform than for the distribution function (in order reduce possible discretisation error). In this scenario, calculation of the IRT would be even simpler.

The IRT arises from the calculation of the directional recoil rate, integrated over a given angular bin, which could then be compared with the number of events observed in that bin. We consider the IRT for two reasons. First, it allows us to perform all of the relevant angular integrals in the calculation of the Radon transform, eliminating the need for computationally-intensive numerical integrals over the angular variables. Secondly, the loss of angular information in discretising the velocity distribution means that the full Radon transform of this discretised distribution is unlikely to provide a good fit to the data on an event-by-event basis. Instead, considering the IRT (and correspondingly binned data) should reduce the error which is introduced by using a discretised approximation to the velocity distribution. This in turn allows us to parametrise the $v$-dependence of each angular bin and mitigate uncertainties in the velocity distribution.

The full details of the derivation of $\hat{f}^j(v_\textrm{min})$ for arbitrary $N$ are included in Appendix.~\ref{app:RadonDeriv}. We include this derivation as a pedagogical tool in the interests of anyone who wishes to modify or extend the approach presented here (for example, by considering different definitions for the bins in $\cos\theta'$ or $\cos\theta$). However, for the reader interested in simply calculating $\hat{f}^j(v_\textrm{min})$, given a set of $f^k(v)$, the algorithm is given in Appendix~\ref{app:Radon}. We note that all of the angular integrals involved can be performed analytically, meaning that computing the discretised Radon transform reduces to performing a series of 1-dimensional integrals over $v$ (one integral for each $\hat{f}^j$), which can be performed numerically. A python code which implements the full calculation of Appendix~\ref{app:Radon} is available from the author.

We note that in the case $N=1$, the IRT simply reduces to an integral over all angles, exactly recovering the velocity integral which appears in non-directional experiments. Also, in the $N=2$ case, the IRT takes the following simple form: 
\begin{align}
\label{eq:directional:N2result}
\hat{f}^1(\vmin) &= 4\pi\int_{\vmin}^{\infty} v \left\{ \pi f^1(v) + \atan\left(\frac{\sqrt{1-\beta^2}}{\beta}\right)\left[f^2(v) - f^1(v)\right] \right\} \, \mathrm{d}v \,,\\
\hat{f}^2(\vmin) &= 4\pi\int_{\vmin}^{\infty} v \left\{ \pi f^2(v) + \atan\left(\frac{\sqrt{1-\beta^2}}{\beta}\right)\left[f^1(v) - f^2(v)\right] \right\} \, \mathrm{d}v\,,
\end{align}
where $\beta = \vmin/v$. We have also checked using Monte Carlo calculations that our calculations give the correct forms for the IRT in the cases $N=2$ and $N=3$. In order to do this, we randomly sample particles from the discretised distributions. For each particle, we simulate a random scattering event and store the values of $\vmin$ and $\hat{\mathbf{q}}$. We then bin these events in angle and compare with the calculated distribution of $\hat{f}^j(\vmin)$. The results are in perfect agreement with the results expected from Appendix~\ref{app:Radon}.

\section{Comparison with exact results}
\label{sec:compare}

We now wish to compare these approximate IRTs with the IRTs obtained from the full (non-discretised) velocity distribution. To do this, we select a benchmark velocity distribution (such as the SHM) and calculate the $f^{k}(v)$ of Eq.~\ref{eq:discretisedf} by averaging over $\cos\theta'$ in each bin, as in Eq.~\ref{eq:averagef} and Fig.~\ref{fig:Discrete}. We then insert these into the algorithm presented in Appendix~\ref{app:Radon} to obtain the corresponding IRTs. We refer to these as the \textit{approximate} IRTs. For comparison, we use the full Radon transform of Eq.~\ref{eq:analRadon} to obtain the \textit{exact} IRTs by integrating over $\cos\theta$. We fix the peak of the underlying speed distribution to be aligned in the forward direction $\theta' = 0^\circ$. However, we discuss the consequences for `misaligned' distributions in Sec.~\ref{sec:discussion}. 

We discuss qualitatively the differences in shape between the exact and approximate IRTs, as any such differences could potentially lead to biases in the reconstruction of particle physics parameters and the velocity distribution. However, directional direct detection experiments do not directly measure the Radon Transform, but rather the differential recoil rate $\mathrm{d}R/\mathrm{d}E_R\mathrm{d}\Omega_q$ of Eq.~\ref{eq:Rate}. Analogously to the previous section, we define the directionally-integrated recoil spectrum:

\begin{equation}
\label{eq:discreteRate}
 \frac{\mathrm{d}R^j}{\mathrm{d}E_R} = \int_{\phi = 0}^{2\pi} \int_{\cos(j\pi/N)}^{\cos((j-1)\pi/N)} \frac{\mathrm{d}R}{\mathrm{d}E_R\mathrm{d}\Omega_q} \, \mathrm{d}\cos\theta \, \mathrm{d}\phi\,.
\end{equation}
The integrated recoil spectrum is then related to the IRT by
\begin{equation}
\label{eq:discreteRate2}
\frac{\mathrm{d}R^j}{\mathrm{d}E_R} \propto  \hat{f}^j(\vmin(E_R)) F^2(E_R)\,,
\end{equation}
where $\vmin(E_R)$ is defined in Eq.~\ref{eq:vmin}. 
The form factor $F^2(E_R)$ leads to a roughly exponential suppression of the rate (relative to the IRT) with increasing $E_R$. Because of this simple proportionality relationship between the IRT and the rate, we do not discuss the shape of the rates in detail.

Instead, we perform a more quantitative comparison between the full and discretised event rates. Assuming a particular idealised detector, we compare the expected number of events  $N_j$ obtained within each bin using the exact and approximate IRTs presented in the previous section: \footnote{Because the $N=1$ case (full average over all angles) is exact, the discretised velocity distribution will always recover the correct total number of signal events. We have checked this explicitly up to $N=10$ by summing the number of events in each angular bin.}
\begin{equation}
\label{eq:Nevents}
N_j = \int_{E_\mathrm{min}}^{E_\mathrm{max}} \frac{\mathrm{d}R^j}{\mathrm{d}E_R} \, \mathrm{d}E_R \,.
\end{equation}
In converting from $\vmin$ to recoil energy $E_R$, it is necessary to fix the values of the DM and nuclear masses. We therefore take as an example a Fluorine target and DM mass of $m_\chi = 50 \,\,\mathrm{ GeV}$. Direct detection experiments do not probe down to arbitrarily low energies and typically an energy threshold is set below which either the signal may be dominated by unrejected backgrounds or the direction of the recoil cannot be reconstructed. We consider a typical threshold energy of 20 keV \cite{Daw:2011}. 

We fix the normalisation of the event rate by requiring a total of 50 signal events across all bins. As previously noted, the anisotropy of a DM signal can be confirmed with around 10 signal events \cite{Morgan:2005}, while the median recoil direction can be confirmed to match that expected due to the Earth's motion with of order 30 events \cite{Green:2010b}. It is reasonable then to expect that the approach presented here might be employed once around 50 signal events are observed. In addition, we include a background of 1 event, distributed isotropically. We discuss the impact of a larger number of signal events in Sec.~\ref{sec:discussion}.

The error between the exact and approximate results can then be compared with the typical Poissonian uncertainty on the expected number of events in each bin $\Delta N_j \sim \sqrt{N_j}$. If the discrepancy between the exact and approximate results is smaller than this Poisson error in the number of events, we would expect the two methods to be in agreement within statistical uncertainties. This means that the data should be approximately as well-fit by the discretised velocity distribution as by the full distribution. This in turn indicates that the discretised velocity distribution can be used, along with a suitable parametrisation, to fit and extract information about the underlying $f(\mathbf{v})$.

We consider two example distributions, which have already been illustrated in Fig.~\ref{fig:Discrete}. The first is the canonical Standard Halo Model (SHM), with parameters $v_{e} = 220 \kms$ and $\sigma_v = 156 \kms$. We use this not only because it is so often studied in the literature, but also because it has a relatively smooth, simple structure and is not too strongly peaked. If the method presented here is to be useful, it should give accurate results for such a simple underlying distribution. For simplicity we do not truncate the SHM at the Galactic escape speed. The inclusion of such a cut-off would reduce the high-speed DM population, reducing the directionality of the signal and therefore improving the results presented here. In the reconstruction of real data, an escape speed cut-off could be included easily in the chosen parametrisation for $f^k(v)$.

As a comparison, we also consider a stream distribution, modeled using Eq.~\ref{eq:SHM_Radon}, but with parameters  $v_{e} = 500 \kms$ and $\sigma_v = 20 \kms$.  This leads to a sharply peaked, strongly directional distribution function, which allows us to compare the SHM with a more extreme case. We note in particular that these parameters lead to a sharper distribution (and therefore a more directional rate) than values typically assumed, for example, for the Sagittarius stream \cite{Freese:2004, Freese:2005, Savage:2006}. This stream should therefore be considered a `worst-case' scenario which is difficult to approximate accurately. 

\subsection{N=2 discretisation}

\begin{figure}[t!]
  \centering
  \includegraphics[width=0.32\textwidth]{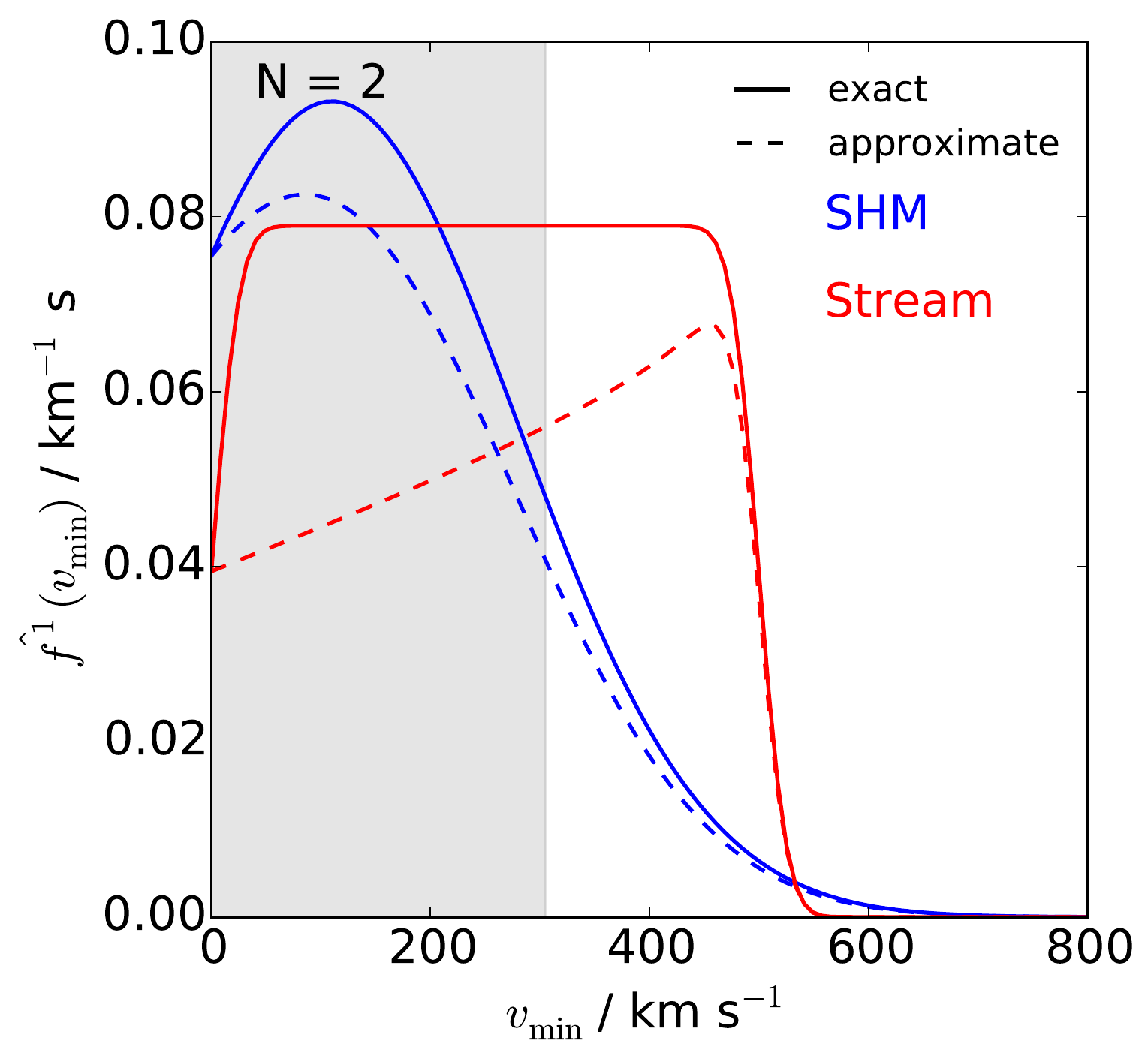}
  \includegraphics[width=0.32\textwidth]{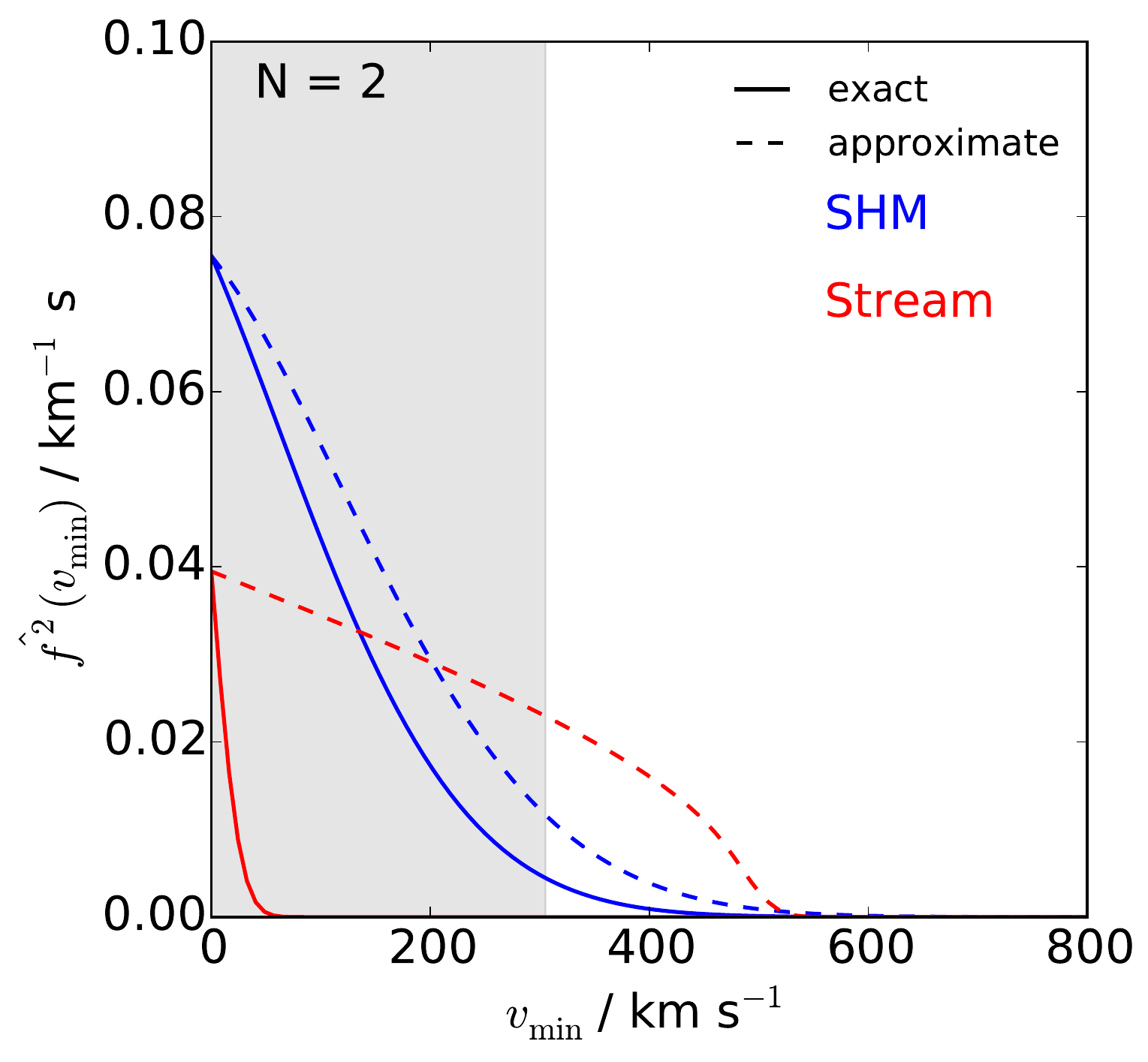}  
\caption{Exact (solid) and approximate (dashed) integrated Radon transforms, $\hat{f}^j$, defined in Eq.~\ref{eq:discreteRadon}, for $N=2$. Results are shown for the SHM (blue) and stream (red) distribution functions. The approximate Radon transforms are obtained by discretising the full velocity distribution into $N$ angular bins. The vector $\textbf{v}_\textrm{e}$ is aligned along $\theta' = 0$. The shaded grey region lies below the energy threshold of 20 keV for a DM mass of 50 GeV.}
  \label{fig:Compare-N=2}
\end{figure}

Figure~\ref{fig:Compare-N=2} shows the comparison between the exact and approximate IRTs for the $N=2$ discretisation. The exact IRT is shown as a solid line, while the approximate IRT obtained from the discretisation is shown as a dashed line. We show the results for both the SHM (blue) and stream distributions (red). Values of $\vmin$ in the shaded region lie below the assumed 20 keV energy threshold for a Fluorine target and 50 GeV DM particle. In the case of the SHM, the shape of the Radon transform is well reconstructed in both the forward (left panel) and backward (right panel) directions, even with the crude $N=2$ discretisation. For the forward IRT ($j = 1$), the ratio between the exact and approximate results is less than $20\%$, while for the backward ($j=2$) case, the error is typically larger ($50-100\%$), though the absolute value is smaller. 

For the stream distribution, the differences between the exact and approximate results are more significant. The results using the discretised velocity distribution underestimate the forward IRT at low $\vmin$, while overestimating the backward IRT for all values of $\vmin$. The reason for this can be seen by examining the velocity distributions illustrated in Fig.~\ref{fig:Discrete}. The full stream distribution (top right) is strongly focused in the forward ($\theta' = 0^\circ$) direction, meaning that particles must scatter through almost $90^\circ$ to produce a recoil in the backward direction. This is kinematically allowed only for a small fraction of the population. Due to the angular averaging, however, the $N=2$ velocity distribution (middle right) has a significant population of DM particles with velocities at right angles ($\theta' = 90^\circ$) to the forward direction. Thus, the discretised velocity distribution has a greater chance of producing scatters in the backward direction. Overall, the discretised distribution is less focused in the forward direction than the full distribution, resulting in a reduced asymmetry between the forward and backward scattering rates.

\begin{figure}[t!]
  \centering
  \includegraphics[width=0.45\textwidth]{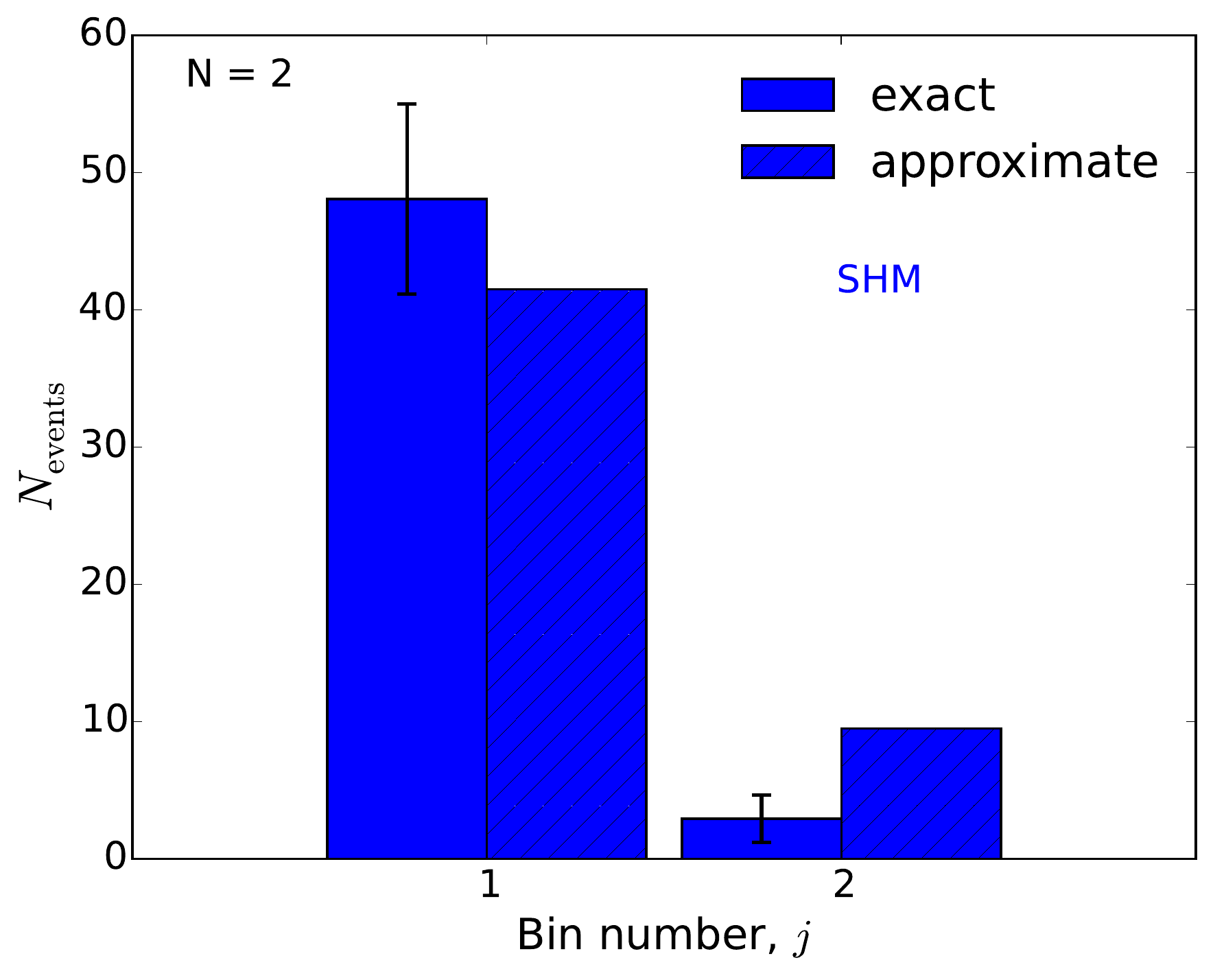}
  \includegraphics[width=0.45\textwidth]{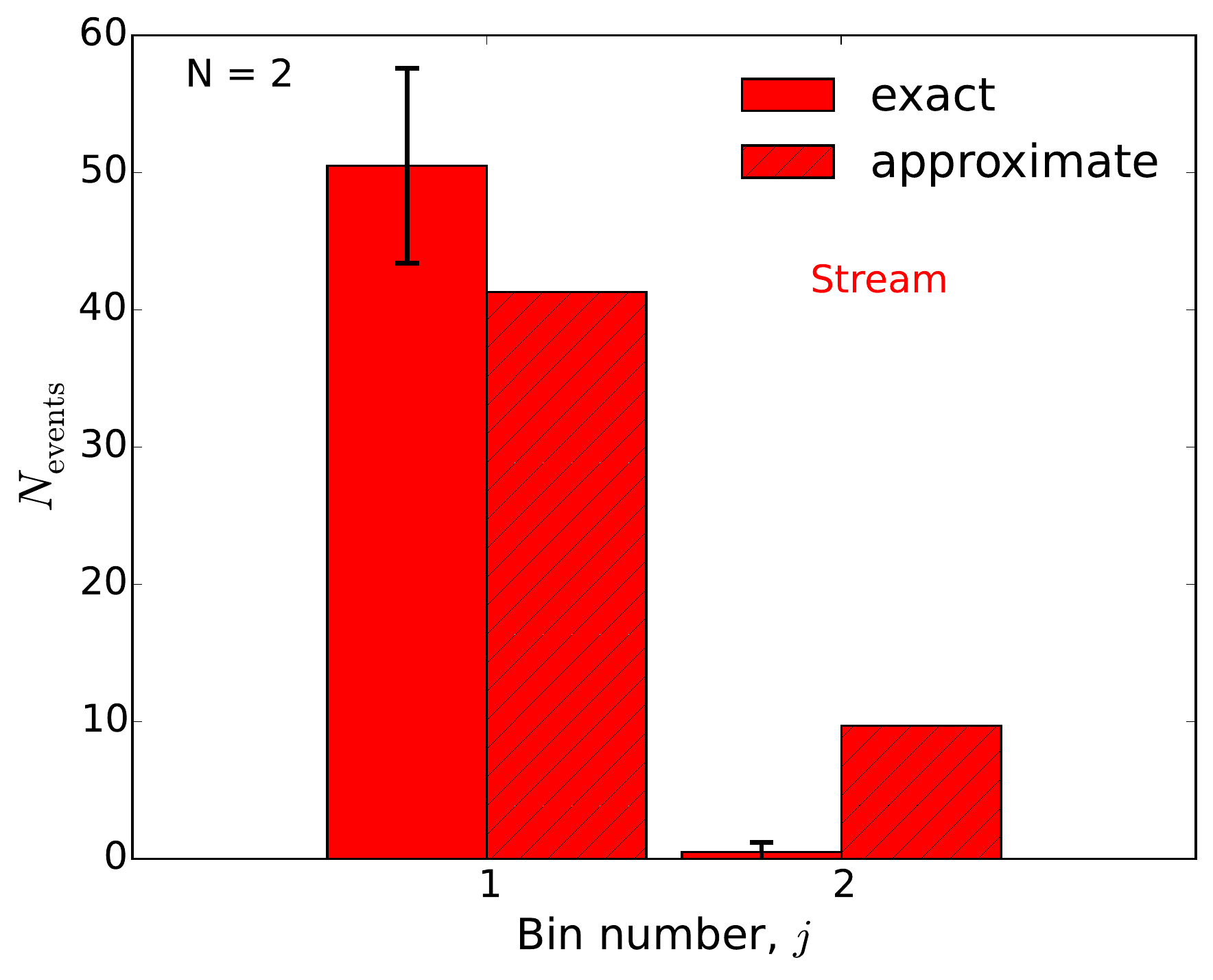}  
\caption{Comparison of the number of events in each angular bin (Eq.~\ref{eq:Nevents}) for N=2 bins, for a Fluorine target with 20 keV threshold and $m_\chi = 50 \,\, \mathrm{GeV}$. Event numbers are calculated using the exact (solid bars) and approximate (hatched bars) IRTs. The SHM distribution is shown in the left panel, while the stream distribution is shown on the right. A total of 50 signal events are assumed (as well as 1 isotropically distributed background event). The error bars on the `exact' number of events are the Poissonian errors $\Delta N_j \sim \sqrt{N_j}$.}
\label{fig:Rate-N=2}
\end{figure}

Figure~\ref{fig:Rate-N=2} shows the comparison between the number of events $N_j$ in each of the angular bins, defined in Eq.~\ref{eq:Nevents}, for the case of $N=2$ discretisation. The exact and approximate calculations are shown as solid and hatched bars respectively, while the SHM and stream distributions are shown separately in the left and right panels. In addition, the `exact' event numbers in each bin are assigned an error bar given by the Poissonian standard deviation $\sqrt{N_j}$. 

In the case of the SHM (left), the number of events in the forward direction ($j=1$) is roughly in agreement (within Poisson uncertainties) when calculated using the exact and approximate IRTs. However, in the backward direction ($j=2$), the approximate calculation overestimates the number of events by a factor of 3. As shown in the left hand panel of Fig.~\ref{fig:Compare-N=2}, in the forward direction, the greatest discrepancy between the IRTs occurs at low $\vmin$. When we consider events only above the energy threshold (i.e.~above the shaded grey region), this discrepancy is therefore minimal. In the right hand panel of Fig.~\ref{fig:Compare-N=2}, however, the fractional discrepancy grows as a function of $\vmin$, and is greatest above the energy threshold of the experiment. This is due to the enhancement in DM particles travelling in the backward direction $\theta' = 180^\circ$ when using the discretised velocity distribution (see Fig.~\ref{fig:Discrete}) and leads to the large discrepancy in the number of backward-going events. 

This problem is even greater for the stream distribution (right), for which neither the forward nor backward event numbers agree within the statistical errors. In the backwards direction, the exact calculation predicts much less than 1 event, while
the approximate calculation overestimates this by a factor of $\sim 30$. For such a highly directional distribution, such a small number of bins is clearly not suitable. 

We note briefly that in the case of $N=2$ discretisation, the approximate calculation of event numbers gives results which are almost indistinguishable for the SHM and stream distributions. This is because the discretised forms of the SHM and stream distributions differ most substantially at low $\vmin$. This can be seen in the middle row of Fig.~\ref{fig:Discrete}; eliminating those particles which are below the threshold speed of $\sim 300 \kms$ gives two similar distributions. Given that there is poor agreement between the number of expected events for the $N=2$ discretisation, and that the discretised SHM and stream distributions are largely indistinguishable, it is clear that such a coarse angular approximation is not suitable for fitting to data.

\subsection{N=3 discretisation}

\begin{figure}[t!]
  \centering
  \includegraphics[width=0.32\textwidth]{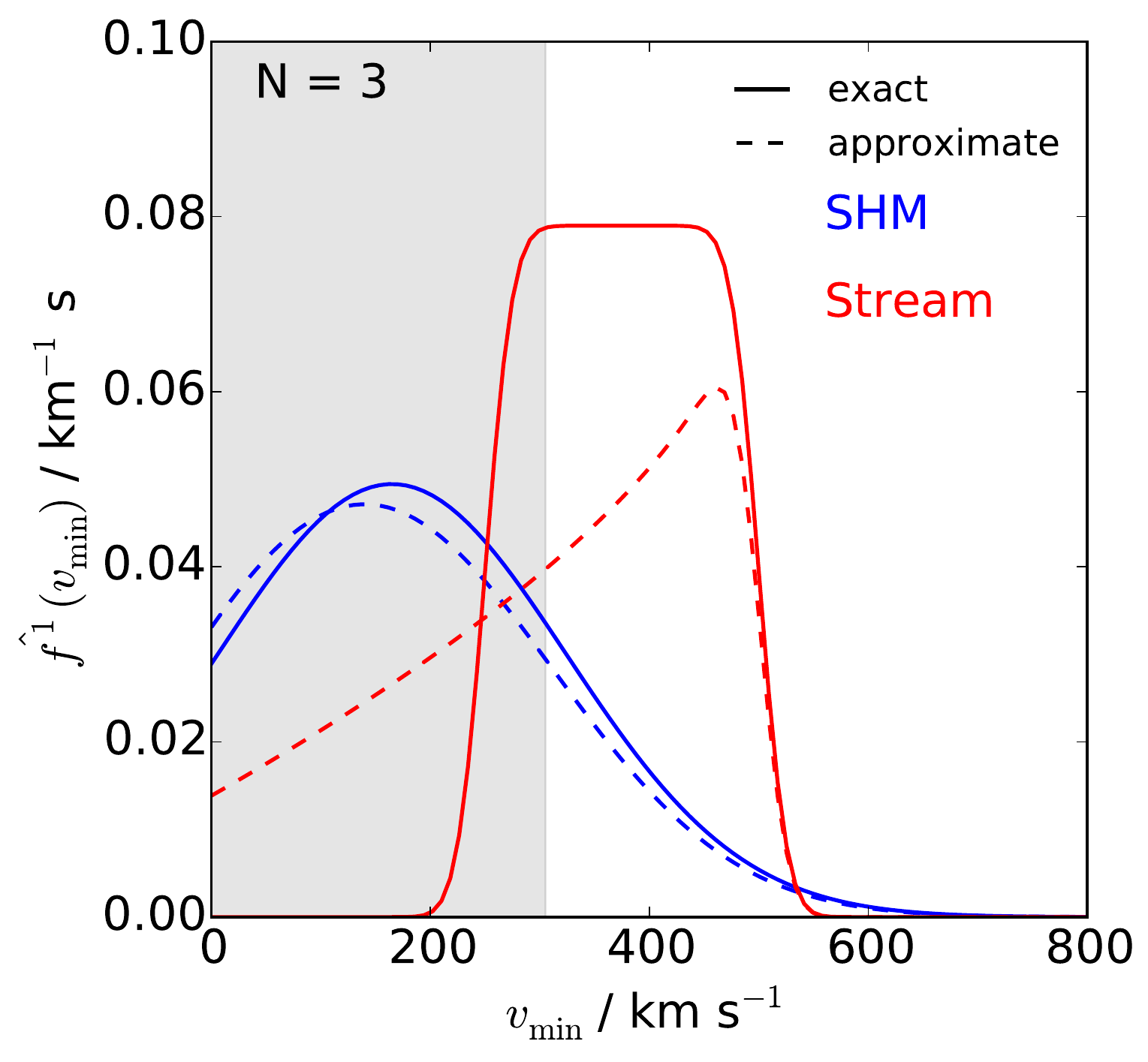}
  \includegraphics[width=0.32\textwidth]{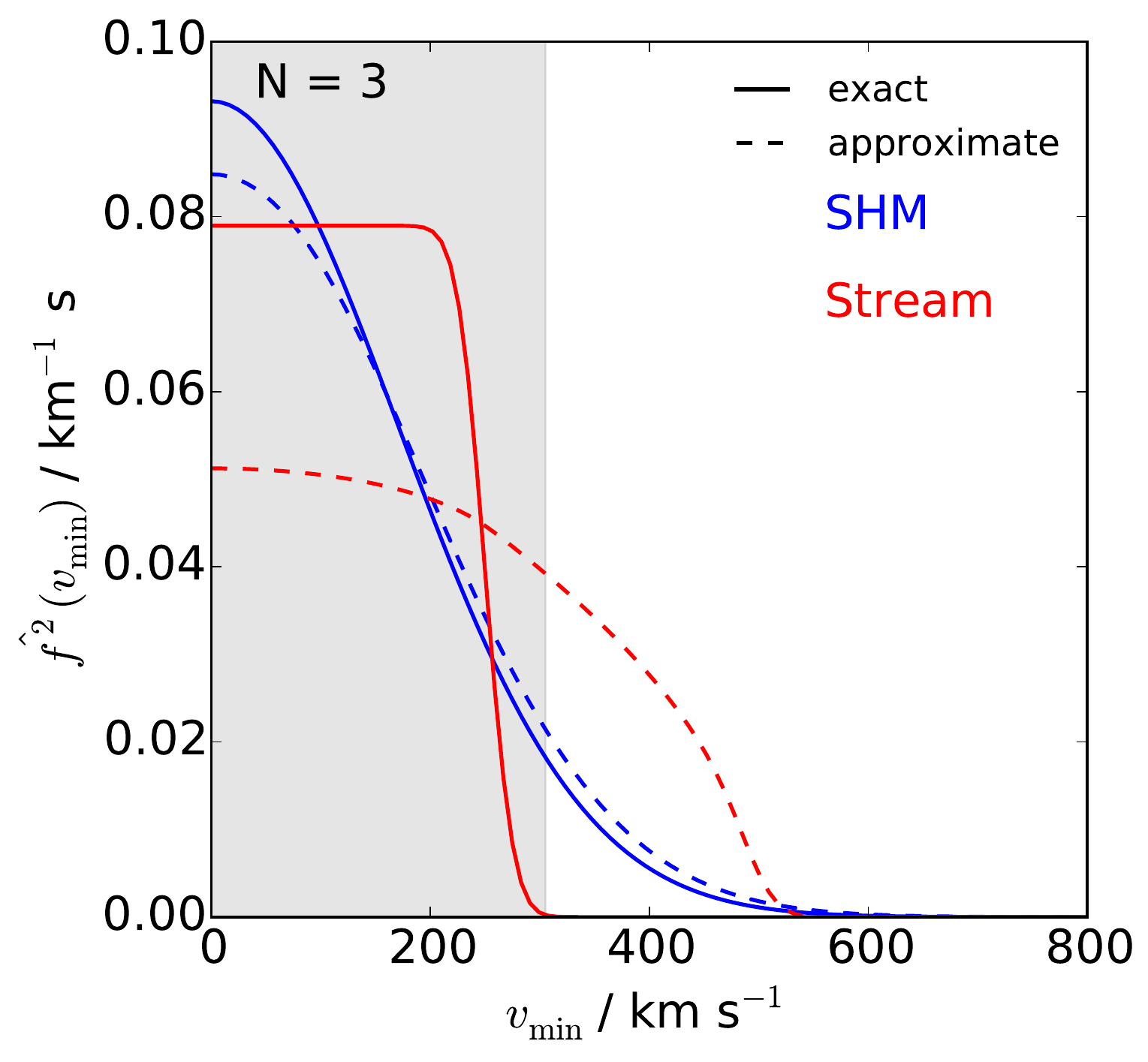}  
  \includegraphics[width=0.32\textwidth]{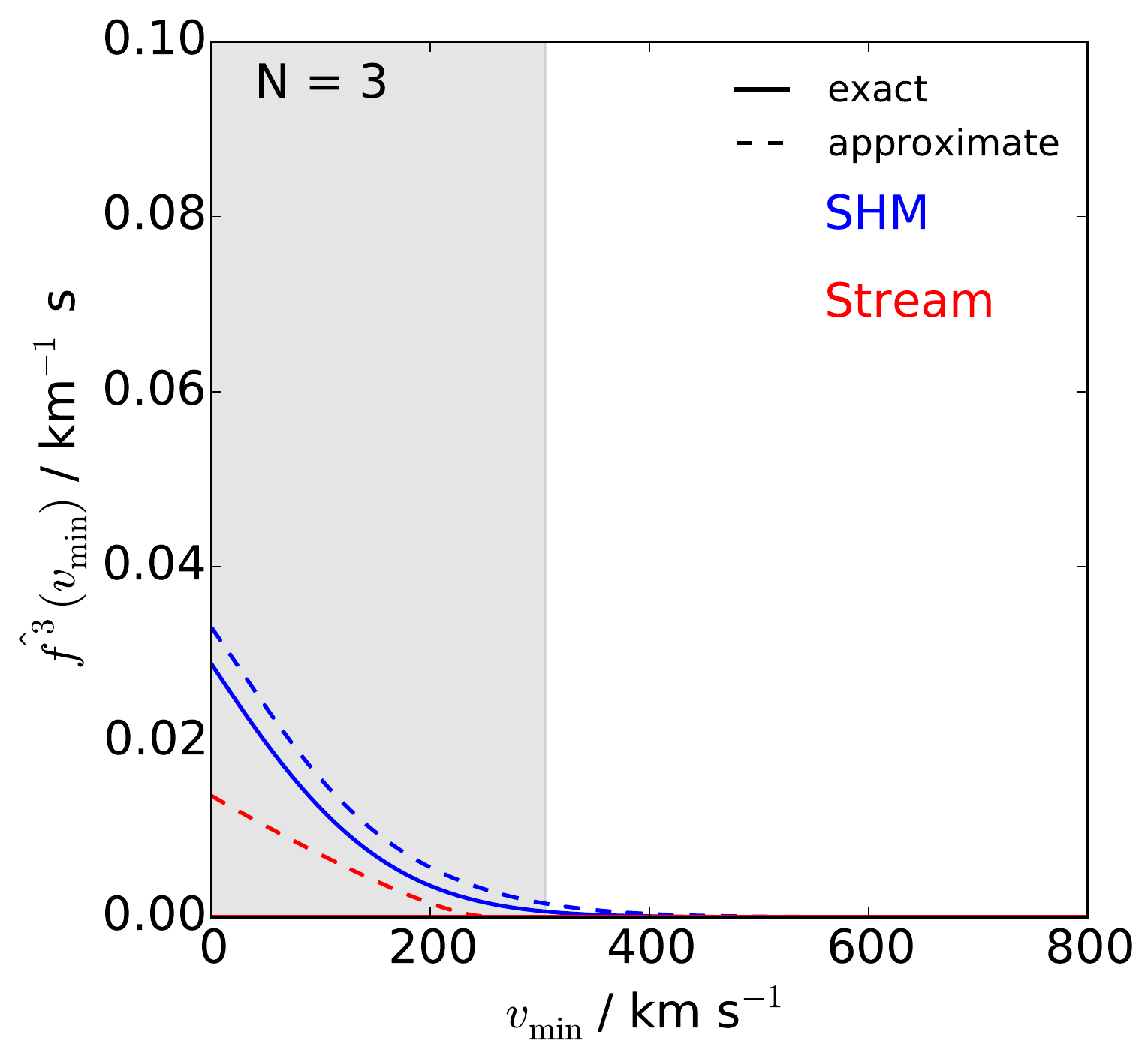}
\caption{As Fig.~\ref{fig:Compare-N=2}, for N = 3 angular bins. In the right panel, the exact Radon transform for the stream is indistinguishable from zero.}
  \label{fig:Compare-N=3}
\end{figure}

Figure~\ref{fig:Compare-N=3} compares the approximate and exact IRTs for the $N=3$ discretisation illustrated on the bottom row of Fig.~\ref{fig:Discrete}. In the case of the SHM, the agreement between the two functions is improved compared to the $N=2$ case. The slope of the event rate is degenerate with the DM mass $m_\chi$, so any error in the shape of the IRTs could translate to a bias in the reconstructed value of $m_\chi$. In the forward direction (left panel), the fractional error is reduced to at most $10\%$ and the shape of the IRT appears to be agree closely in all three directional bins. This suggests that any such bias should be minimal. 

However, there is still a significant difference between the shapes of the exact and approximate IRTs for the stream distribution. For example, in the forward direction (left panel), the exact calculation predicts scattering events occurring only with a narrow range of $\vmin$, from $250 \kms$ to $500 \kms$. This is because almost all particles are moving in the forward direction with speed $v = 500 \kms$. The kinematics of the scattering requires that $\mathbf{v}\cdot\hat{\mathbf{q}} = \vmin$, meaning that scattering in the forward direction ($\theta = 0^\circ$) is only allowed for $\vmin \approx 500 \kms$. Scattering away from the forward direction, but still included in the first angular bin ($\theta < 60^\circ$), is kinematically allowed for $\vmin > \cos(60^\circ) \times 500 \kms  \approx 250 \kms$. For the discretised velocity distribution, there is a population of DM particles initially directed away from the forward direction ($\theta \neq 0^\circ$) which can scatter through larger angles and still produce scattering events within the first angular bin. These large-angle scattering events mean that values of $\vmin$ down to zero are now kinematically allowed. 

Considering the $j=3$ angular bin (right panel), we note that the exact stream distribution predicts no scattering events in the backward direction. This is because the particles would have to scatter through an angle much larger than $90^\circ$ to produce events in the $j=3$ bin, which is not kinematically allowed. However, for the discretised distribution (bottom right panel of Fig.~\ref{fig:Discrete}), it is still kinematically possible for particles on the edge of the forward bin ($\theta' \approx 60^\circ$) to scatter into the backward direction ($\theta > 120^\circ$). Thus, the approximate calculation still predicts a small but non-zero IRT for the $j=3$ case.

\begin{figure}[t!]
  \centering
  \includegraphics[width=0.45\textwidth]{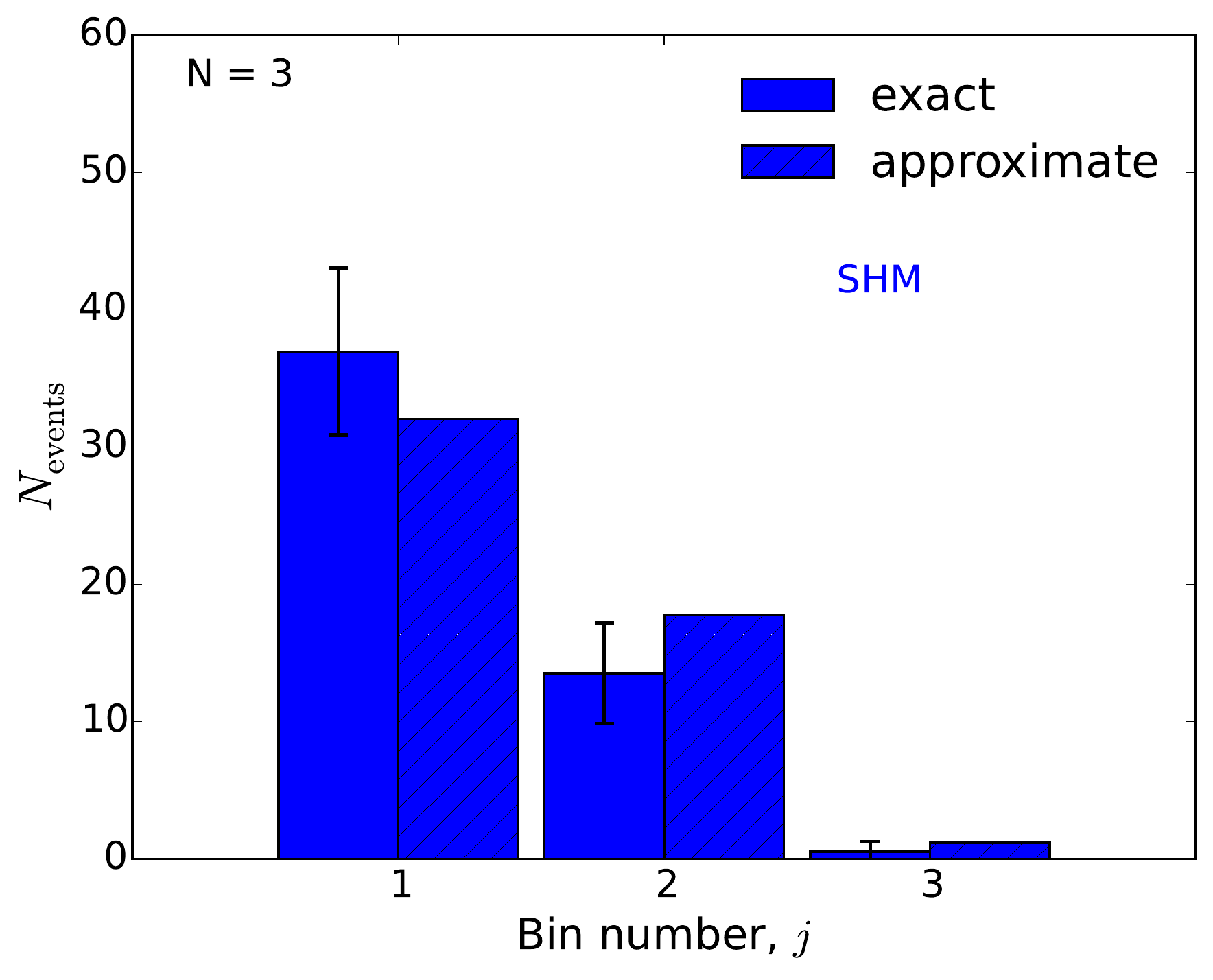}
  \includegraphics[width=0.45\textwidth]{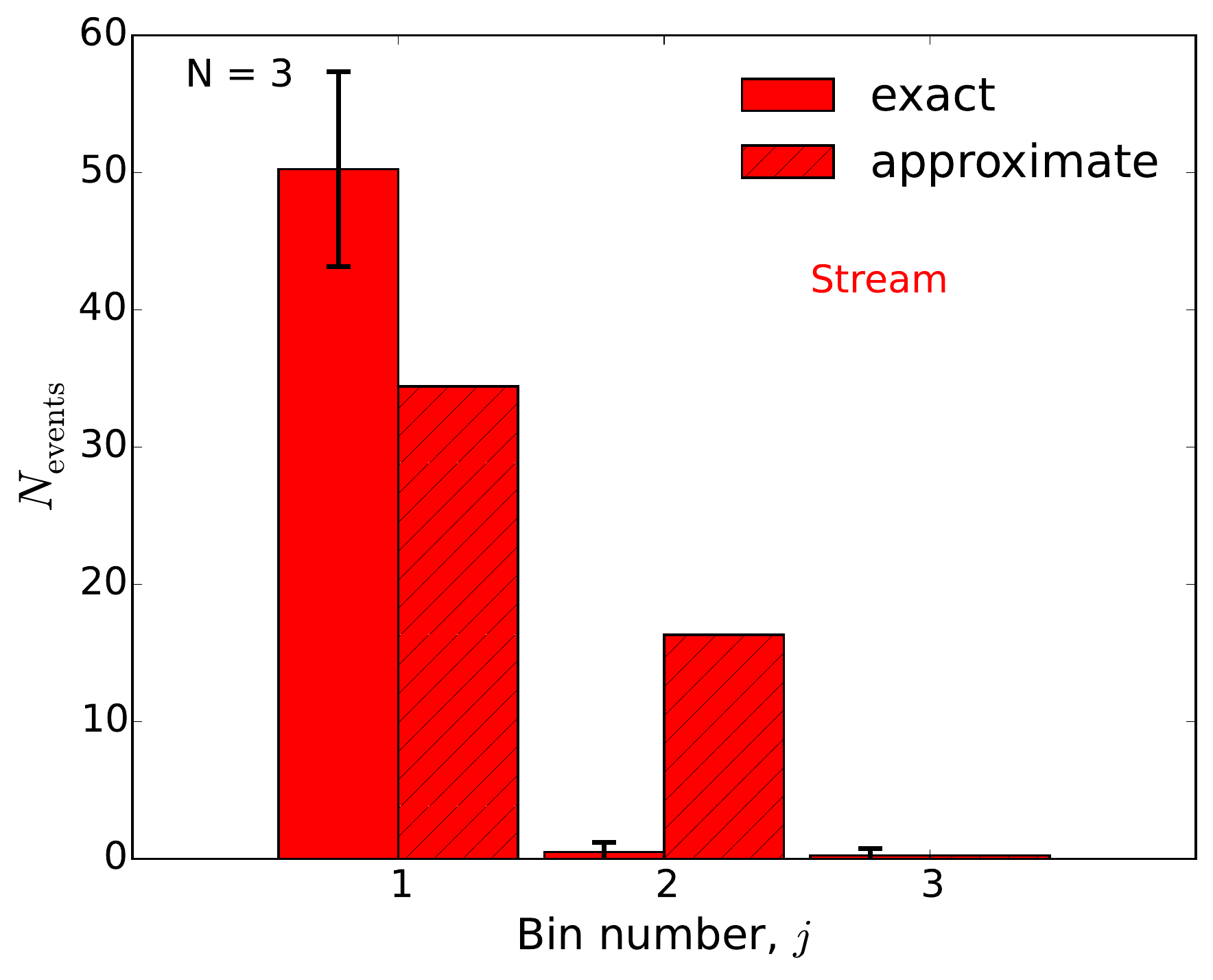}  
\caption{As Fig.~\ref{fig:Rate-N=2}, for N = 3 angular bins.}
\label{fig:Rate-N=3}
\end{figure}

Figure~\ref{fig:Rate-N=3} shows the event numbers expected in each of the $N = 3$ angular bins calculated using the exact and approximate IRTs. For the SHM (left panel), in the forward direction ($j=1$), the approximate calculation underestimates the rate by around 13\%, while in the transverse direction ($j=2$), the approximate calculation overestimates the rate by around 30\%. Finally, in the backwards direction ($j=3$) the rate is overestimated by a factor of 2. However, these discrepancies are roughly the same magnitude as the typical Poisson error in each bin for a sample of 50 events. In contrast to the $N=2$ case then, the $N=3$ case can provide a reasonable approximation to the event rate for the SHM.

For the stream distribution (right panel), there are significant discrepancies, which remain larger than the typical statistical error. In particular, there is a substantial leakage of events from the forward direction ($j=1$) into the transverse direction ($j=2$) in the approximate calculation. This can be seen in the middle panel of Fig.~\ref{fig:Compare-N=3}, in which the exact IRT is zero above the experimental threshold, while the approximate IRT is substantially non-zero. This in turn arises due to the `smearing' of the distribution away from the forward direction, as previously discussed. In contrast, in the backward direction ($j=3$), the approximate calculation now agrees with the exact calculation, giving zero events. As shown in the right panel of Fig.~\ref{fig:Rate-N=3}, the discrepancy between the exact and approximate IRTs arises only at low $\vmin$ (below the energy threshold of the experiment), where scattering into the backwards bin is kinematically allowed.

\subsubsection{The folded rate}

The $N=3$ case is important for the scenario where head-tail discrimination is not possible when reconstructing recoil tracks. Head-tail discrimination of tracks has previously been demonstrated \cite{Burgos:2008}, but may not be possible with 100\% accuracy in near-future detectors \cite{Billard:2012}. Within the formalism considered here, detectors without sense discrimination cannot distinguish between a recoil with direction $\cos\theta$ and another with direction $-\cos\theta$. They therefore probe the so-called `folded' recoil spectrum:

\begin{equation}
\label{eq:folded}
\frac{\mathrm{d}R}{\mathrm{d}E_R\mathrm{d}|\cos\theta|} = \frac{\mathrm{d}R}{\mathrm{d}E_R\mathrm{d}\cos\theta} + \frac{\mathrm{d}R}{\mathrm{d}E_R\mathrm{d}(-\cos\theta)}\,.
\end{equation}

For the $N=2$ case, the folded spectrum is entirely isotropic, as the forward and backward IRTs differ only in the sign of $\cos\theta$. However, in the $N=3$ case, the transverse IRT, given by

\begin{equation}
\label{eq:transverse}
\hat{f}^T(\vmin) = \hat{f}^2(\vmin) = \int_{\phi = 0}^{2\pi}  \int_{-1/2}^{1/2} \hat{f}(\vmin, \cos\theta) \,\mathrm{d}\cos\theta \, \mathrm{d}\phi\,,
\end{equation}
is invariant under $\hat{f}(\vmin, \cos\theta) \rightarrow \hat{f}(\vmin, -\cos\theta)$. That is, the transverse event rate `folds' back onto itself. Thus, even without sense discrimination, directional experiments will still be sensitive to this transverse scattering rate. By comparison, if the forward and backward directions cannot be distinguished, the remaining two IRTs (the left and right panels in Fig.~\ref{fig:Compare-N=3}) are folded together, to obtain the longitudinal Radon transform
\begin{equation}
\label{eq:longitudinal}
\hat{f}^L(\vmin) = \int_{\phi = 0}^{2\pi}\int_{-1}^{-1/2} \hat{f}(\vmin, \cos\theta, \phi) \,\mathrm{d}\cos\theta \, \mathrm{d}\phi + \int_{\phi = 0}^{2\pi}\int_{1/2}^{1} \hat{f}(\vmin, \cos\theta) \,\mathrm{d}\cos\theta \, \mathrm{d}\phi\,.
\end{equation}
Analogously, we can define the longitudinal and transverse event spectra, \dbd{R^{L,T}}{E_R}, and event numbers $N^{L,T}$, derived from the corresponding Radon transforms.

\begin{figure}[t!]
  \centering
  \includegraphics[width=0.45\textwidth]{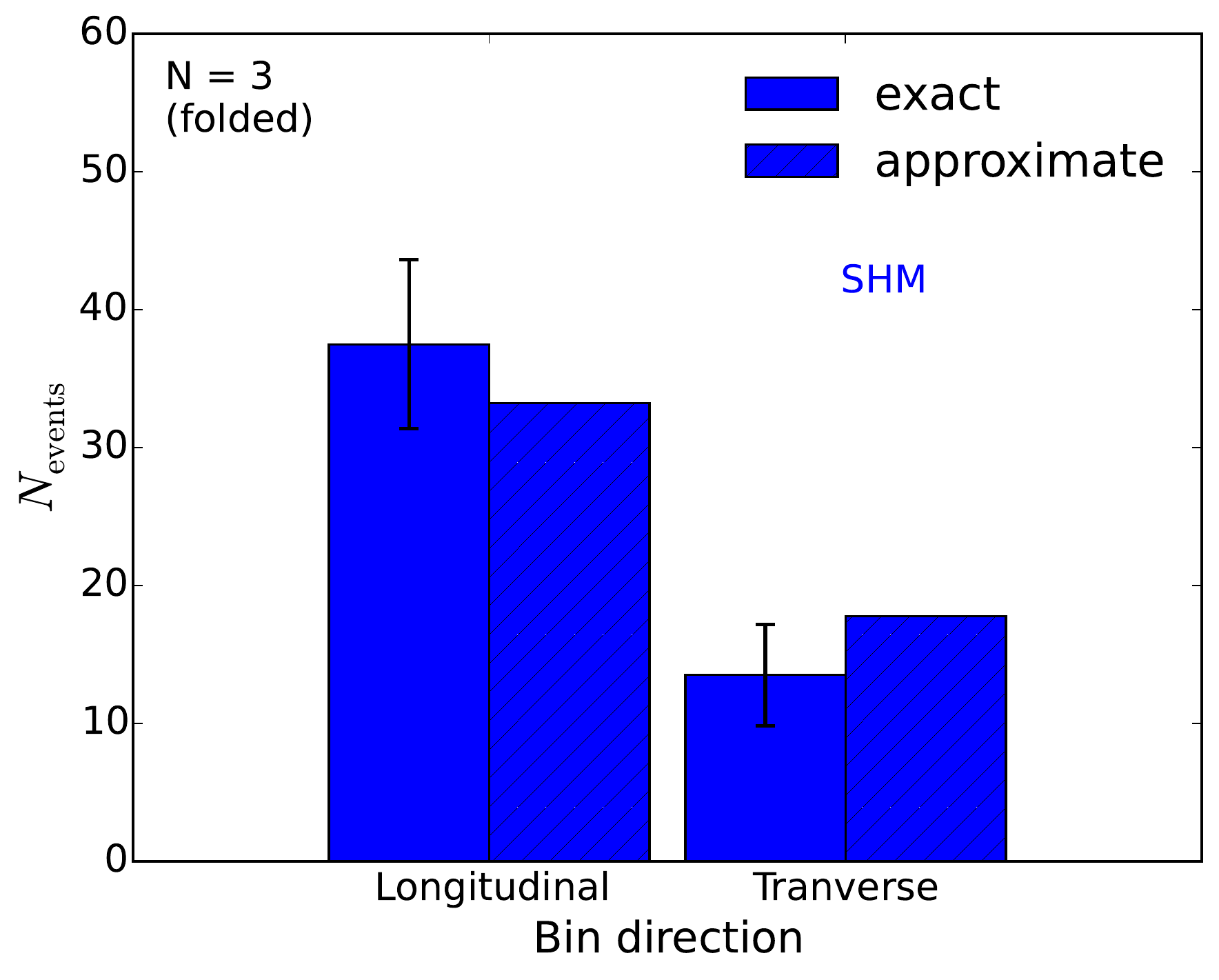}
  \includegraphics[width=0.45\textwidth]{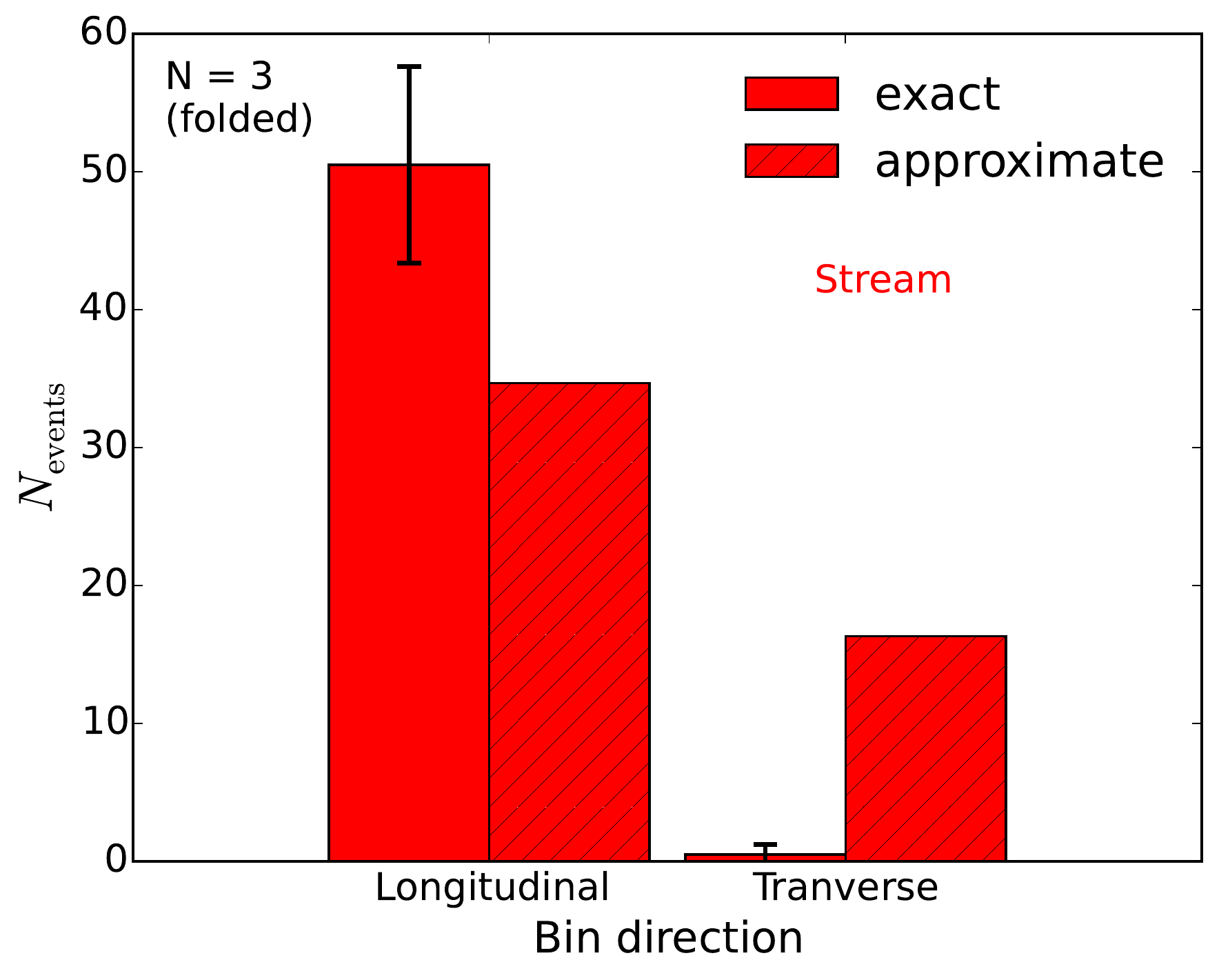}  
\caption{As Fig.~\ref{fig:Rate-N=2}, for the folded event rate with $N=3$ underlying bins, producing one longitudinal and one transverse directional bin (see Eq.~\ref{eq:folded} and associated text).}
\label{fig:Rate-N=3_folded}
\end{figure}

Figure~\ref{fig:Rate-N=3_folded} shows the event numbers for this folded rate in the longitudinal and transverse bins. For the SHM (left panel), there is a slight improvement in the agreement between the exact and approximate results compared with Fig.~\ref{fig:Rate-N=3}. The backward bin, which was previously overestimated by a factor of 2, has been removed and merged with the forward bin to form the longitudinal bin. Because the approximate result previously underestimated the true number of events in the forward bin, this combination of the bins leads to marginally improved agreement, though the effect is small in this example.

In the case of the stream (right panel), the number of events in each bin appears almost indistinguishable from those of Fig.~\ref{fig:Rate-N=3}. This is because the leakage of the events into the backwards direction when using the approximate distribution is minimal and therefore the effect of folding is also minimal. However, we have demonstrated that this method can still be employed when sense discrimination of recoils is not possible and, in general, it should lead to a reduction in the associated discretisation error.

\subsection{N=5 discretisation}

\begin{figure}[t!]
  \centering
  \includegraphics[width=0.32\textwidth]{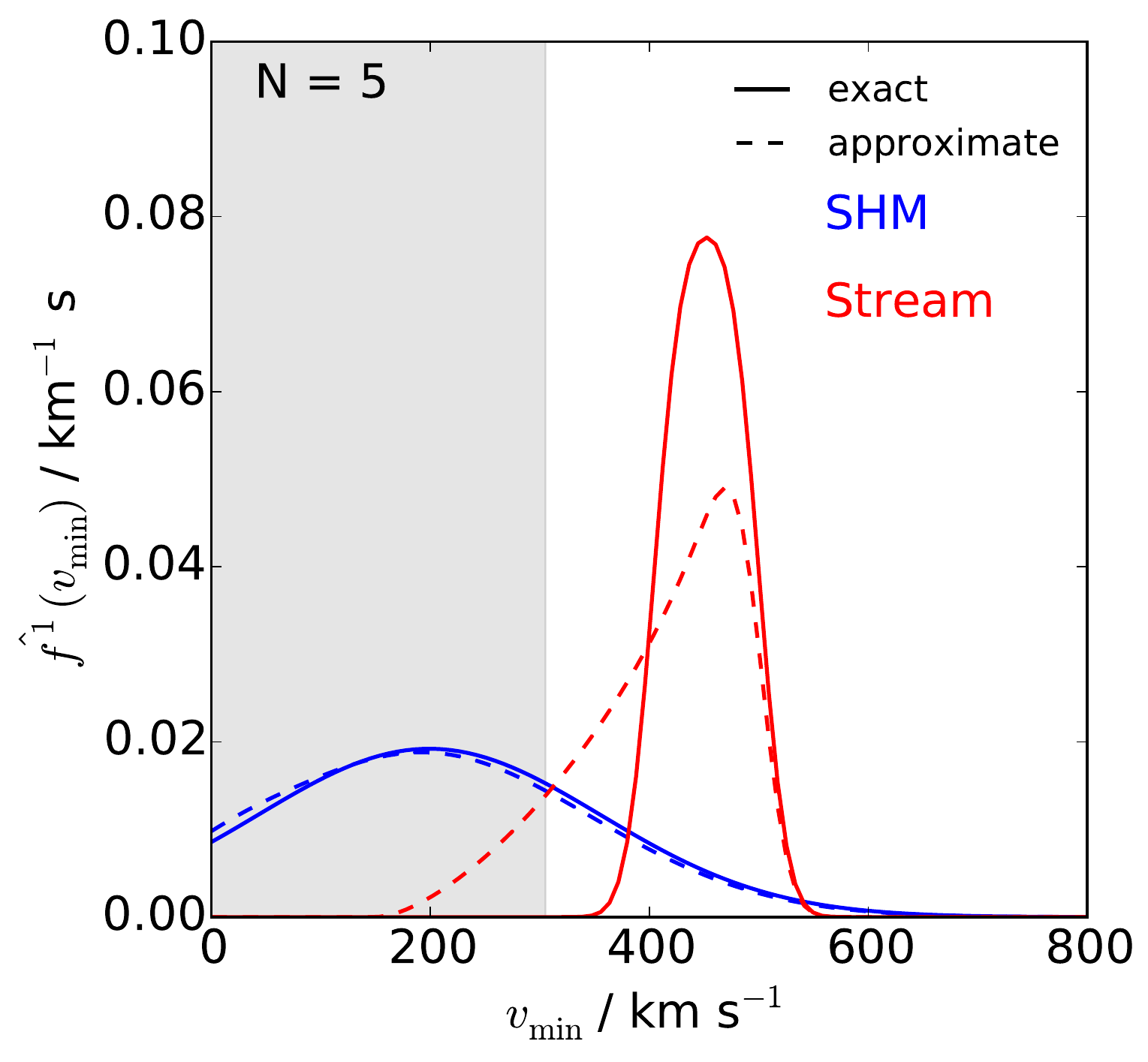}
  \includegraphics[width=0.32\textwidth]{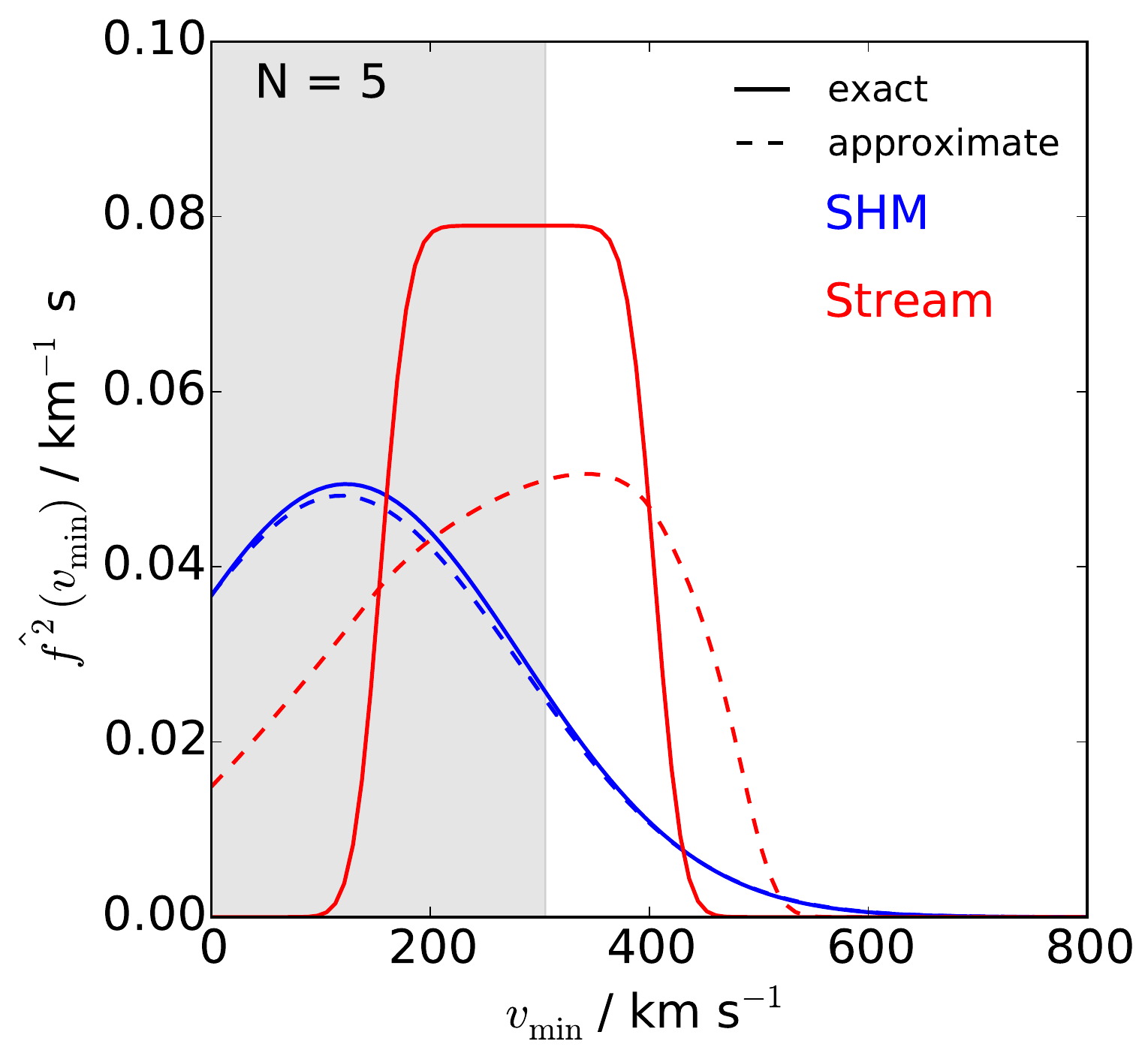}  
  \includegraphics[width=0.32\textwidth]{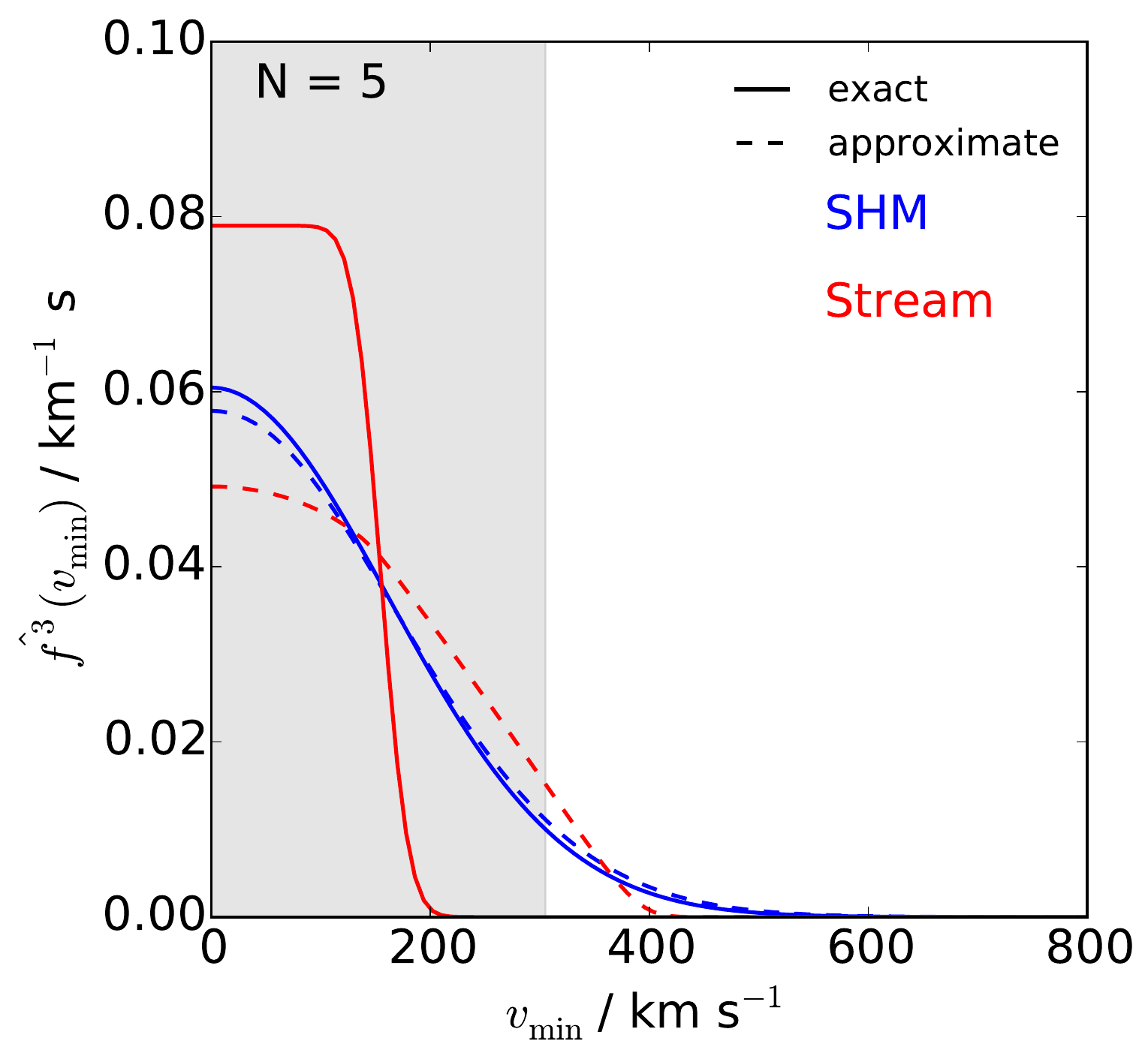}
  \includegraphics[width=0.32\textwidth]{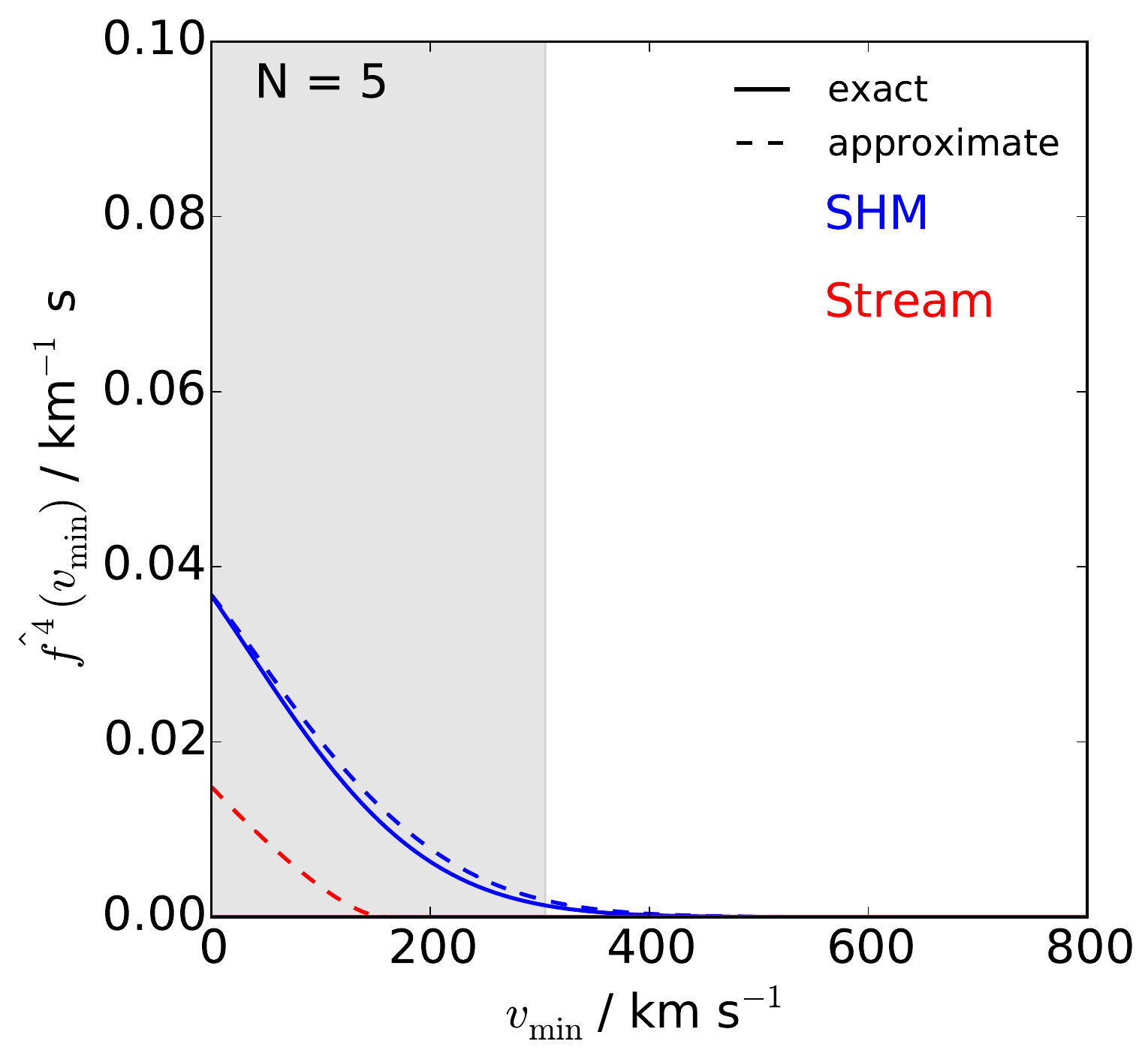}  
  \includegraphics[width=0.32\textwidth]{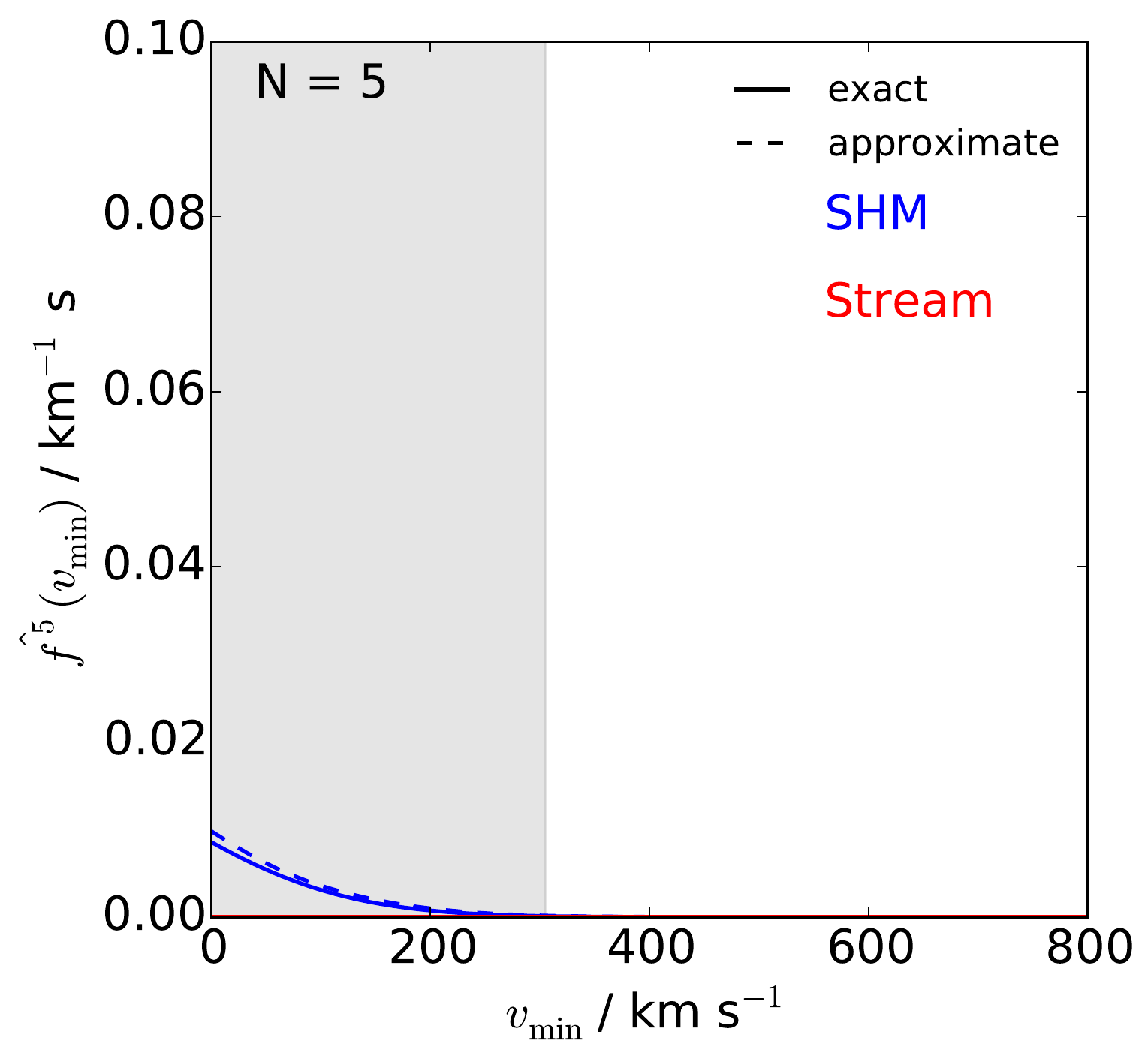}
\caption{Exact (solid) and approximate (dashed) integrated Radon transforms, $\hat{f}^j$ defined in Eq.~\ref{eq:discreteRadon}, for $N=5$. Results are shown for the SHM (blue) and stream (red) distribution functions. The approximate Radon transforms are obtained by discretising the full velocity distribution into $N$ angular bins. The vector $\textbf{v}_\textrm{e}$ is aligned along $\theta' = 0$. In the lower left panel, the exact IRT for the stream is indistinguishable from zero, while in the lower right panel, both the approximate and exact IRTs are indistinguishable from zero.}
  \label{fig:Compare-N=5}
\end{figure}

We now consider a more finely discretised velocity distribution, namely $N=5$. The comparison of the exact and approximate IRTs is shown in Fig.~\ref{fig:Compare-N=5}. The results for the SHM, showing typical discrepancies at the 5\% level, indicate that the approximate Radon transform does in fact tend to the true Radon transform in the limit of large $N$. The results for the stream again show much poorer agreement. We expect that the agreement will not improve significantly until the angular size of each bin is close to the angular extent of the underlying distribution function.
For the stream, we can see by eye in Fig.~\ref{fig:Discrete} that most of the distribution is distributed over an angle $\theta' \lesssim 10^\circ$, meaning that roughly $N=18$ bins are required to prevent the smearing of the distribution function visible for $N=2$ and $N=3$. In spite of this, some structures (such as the peak in the forward direction of the upper left panel) are better reconstructed than in the $N=3$ case, indicating that additional information is still gained by adding more bins. In principle, there is no obstacle to increasing the number of bins up to (and beyond) $N=18$, as the formalism of Appendix~\ref{app:Radon} is applicable for arbitrary $N$. However, the appropriate number of bins would typically be dictated by the amount of data available, with a large amount of data required to justify fitting the number of parameters associated with $N=18$ bins. 

We note that for the bin focused in the backward direction, $j=5$ (bottom right), the exact and approximate IRTs for the stream are indistinguishable and very close to zero. As in the $N=3$ case, scattering in the backward direction is not possible for the full stream distribution because of the strong directionality. For the discretised distribution, most of the distribution is focused in the forward-directed bin ($k=1$ in Eq.~\ref{eq:averagef}). In contrast to the $N=3$ result, however, this is sufficiently directional that scattering into the backward direction (into the $j=5$ bin) is no longer kinematically allowed, resulting in significantly closer agreement in the backward direction.

\begin{figure}[t!]
  \centering
  \includegraphics[width=0.45\textwidth]{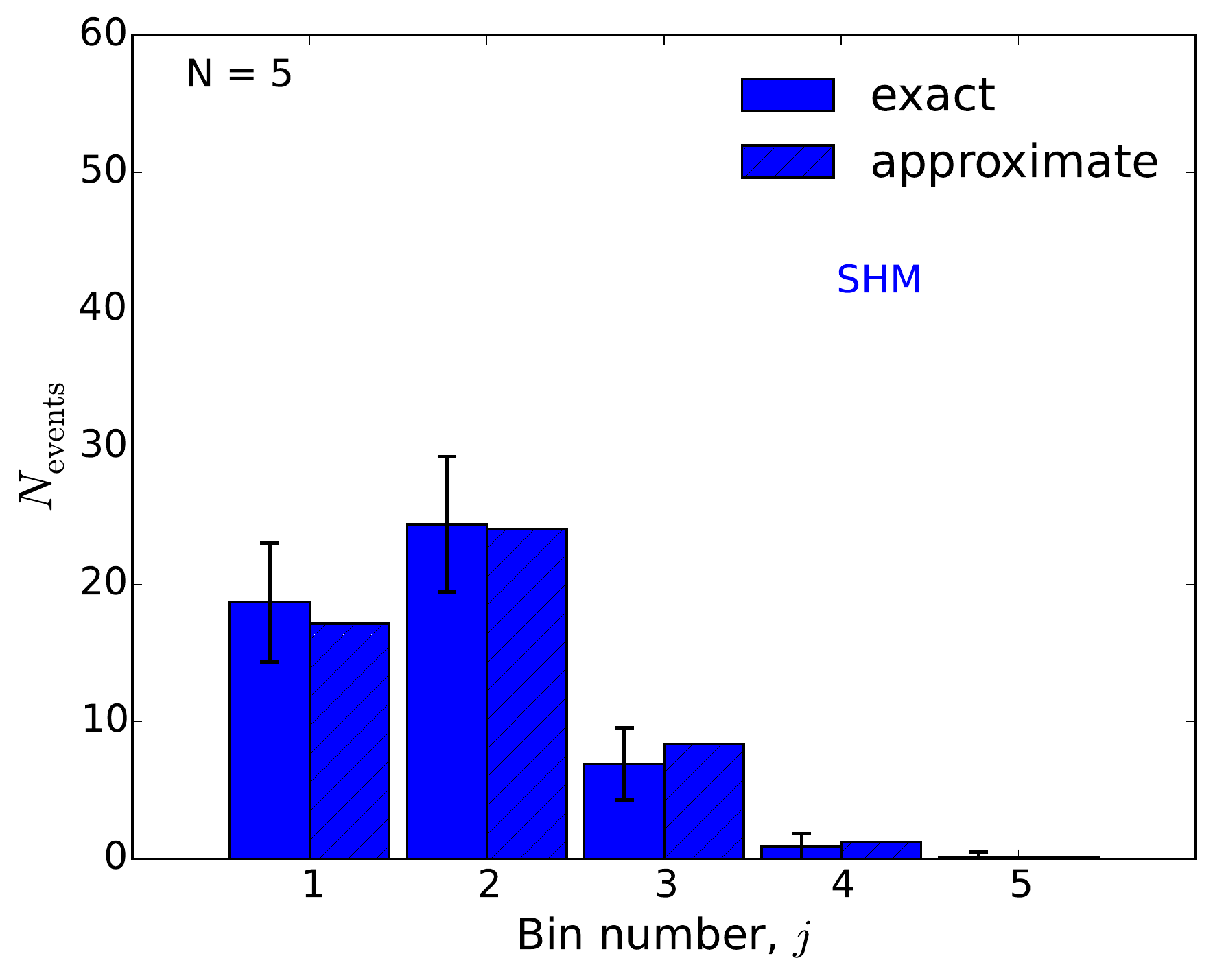}
  \includegraphics[width=0.45\textwidth]{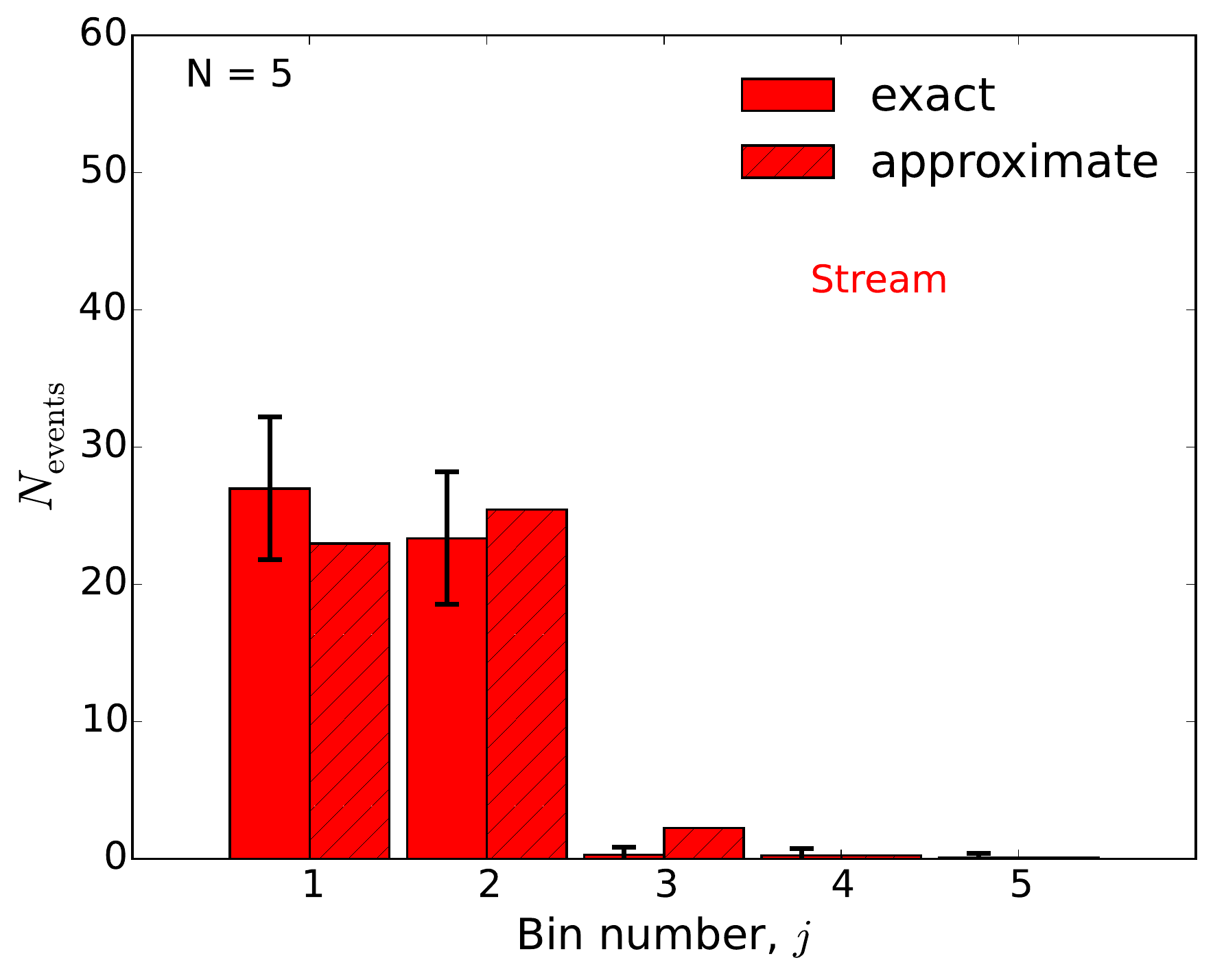}  
\caption{As Fig.~\ref{fig:Rate-N=2}, for N=5 angular bins.}
\label{fig:Rate-N=5}
\end{figure}

In fig.~\ref{fig:Rate-N=5}, we show the event numbers in each bin for the $N=5$ discretisation. For the SHM (left panel), there is now close agreement between the exact and approximate results, with less than 10\% discrepancy for the forward bins $j=1,2$ and an error between 20\% and 40\% in the transverse and backward bins $j=3,4,5$. For all bins, the agreement between the exact and approximate results is significantly smaller than the Poisson uncertainty for the example of 50 signal events. This closely matches the expectation from Fig.~\ref{fig:Compare-N=5}, which shows close agreement between the exact and approximate IRTs. 

For the stream (right panel), there is an improved fit between the exact and approximate results, with bins $j=1,2,4,5$ showing agreement within the statistical uncertainty. Leakage of events between the two forward bins leads to an slight overestimation of the $j=2$ rate. However, the biggest problem remains leakage of events into the transverse bin $j=3$ which gives a significant overestimation of the number of events when using the approximate IRT. Once again, this is seen easily in the upper right panel of Fig.~\ref{fig:Rate-N=5}, where the approximate IRT is non-zero above the energy threshold of the experiment.  

We note that while there remains a notable discretisation error for the stream distribution, it is now possible to distinguish the results for the SHM and stream distribution when using the approximate IRTs, in contrast to $N=2$ discretisation. For example, in the $j=3$ bin, the approximate calculation predicts around $\sim8$ events for the SHM, compared to $\sim 2$ events for the stream. This difference is larger than the error induced by the discretisation (a discrepancy of $\sim 2$ events for both the SHM and stream). This indicates that while the $N=5$ discretisation may not be able to accurately reproduce the event rate for a stream distribution, it may still be useful for distinguishing between different distributions. For example, if the $N=5$ discretisation were used to fit to data and the results indicated a smaller number of events than expected in the $j=3$ bin, this could point to deviations from the SHM. 

\subsection{Forward-backward asymmetry}

Finally, we calculate the forward-backward asymmetry in the number of events, $A_\mathrm{FB}$, and compare the exact and approximate results. In terms of the number of forward and backward events between the threshold and maximum energies $E_\mathrm{th}$ and $E_\mathrm{max}$,

\begin{align}
N_\mathrm{F} &=  \int_{E_{\mathrm{th}}}^{E_\mathrm{max}} \int_{\phi = 0}^{2\pi} \int_{0}^{1} \frac{\mathrm{d}R}{\mathrm{d}E_R\mathrm{d}\Omega_q} \, \mathrm{d}\cos\theta\, \mathrm{d}\phi \, \mathrm{d}E_R \\
N_\mathrm{B} &= \int_{E_{\mathrm{th}}}^{E_\mathrm{max}} \int_{\phi = 0}^{2\pi} \int_{-1}^{0} \frac{\mathrm{d}R}{\mathrm{d}E_R\mathrm{d}\Omega_q} \, \mathrm{d}\cos\theta\, \mathrm{d}\phi \, \mathrm{d}E_R\,,
\end{align} 
the asymmetry is given by

\begin{equation}
\label{eq:AFB}
A_\mathrm{FB} = \frac{N_\mathrm{F} - N_\mathrm{B}}{N_\mathrm{F} + N_\mathrm{B}}\,.
\end{equation}

By comparing the values of $A_\mathrm{FB}$ obtained from the full and discretised distributions, we can obtain a measure of how well the discretised distribution captures the directionality of the signal, as well as showing how this improves with increasing $N$. In addition, this forward-backward asymmetry would potentially be the first directional signal measured with directional detectors and so determining how well it can be recovered is of substantial importance. Using the current formalism, we can consider only even values of $N$, for which each angular bin lies entirely in either the forward or backward direction. In principle, the calculation can also be extended to include odd values of $N$ using the framework of Appendix~\ref{app:RadonDeriv}.

Figure~\ref{fig:RateComparison} shows the forward-backward asymmetry obtained from the discretised velocity distribution for $N=2,4,6,8,10$ (filled squares), compared to the true forward-backward asymmetry, obtained from the full velocity distribution (solid lines). In the case of the stream (red), the event rate is strongly asymmetric ($A_{\mathrm{FB}} \approx 1$) as the velocity distribution is highly focused in the forward direction.  By comparison, the SHM (blue) has a smaller asymmetry  ($A_{\mathrm{FB}} \approx 0.9$) due to its wider velocity dispersion. Dotted lines show the $1\sigma$ statistical uncertainties on the measured value of $A_\mathrm{FB}$ assuming 50 signal events.

\begin{figure}[t!]
  \centering
  \includegraphics[width=0.5\textwidth]{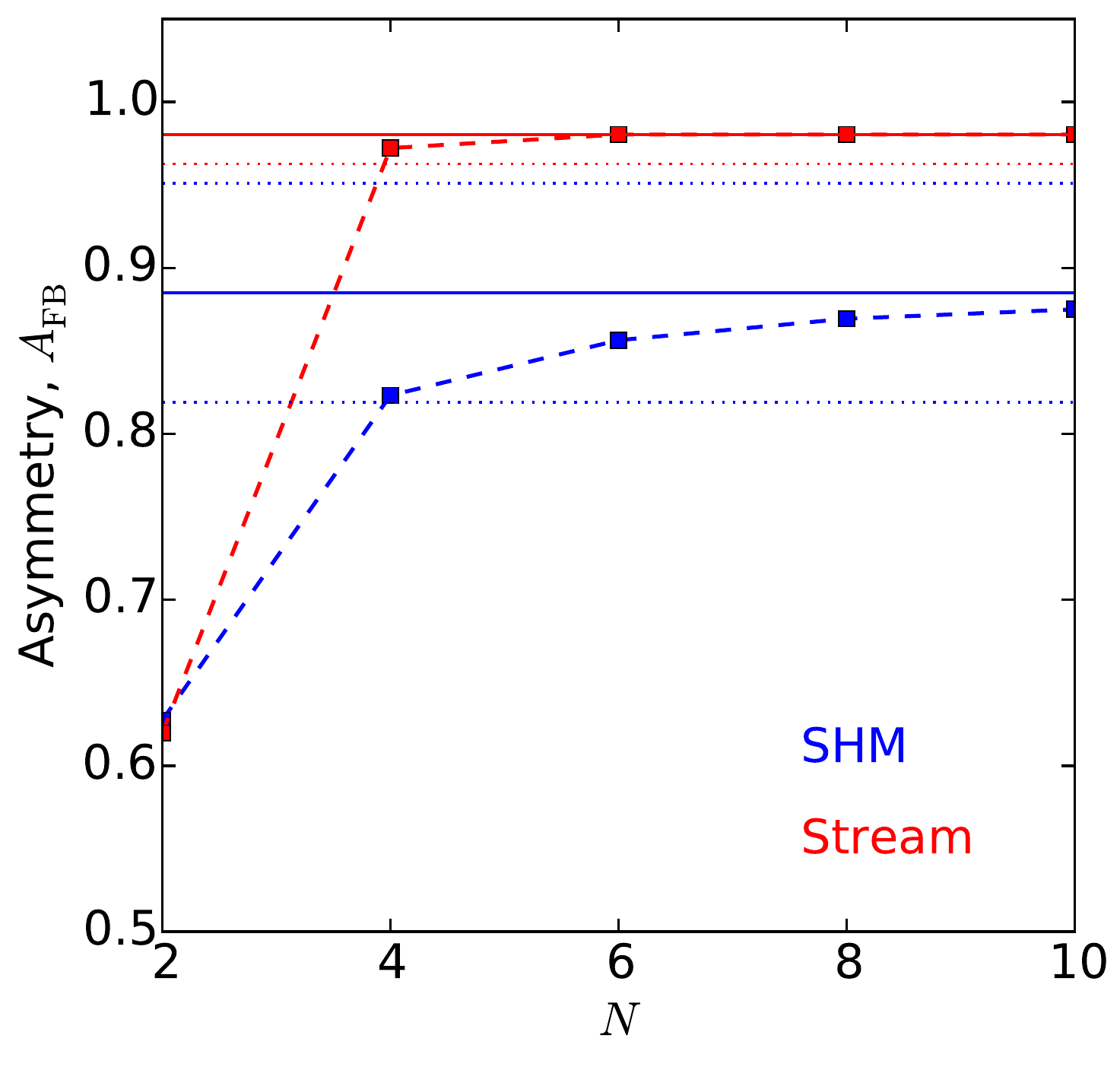}
\caption{Forward-backward asymmetry of the event rate $A_{\mathrm{FB}}$, defined in Eq.~\ref{eq:AFB}, using the full velocity distribution (solid lines) and the discretised velocity distribution (filled squares). Results are shown for the SHM and stream distributions in blue and red respectively. Dotted lines show the $1\sigma$ statistical uncertainties on the measured value of $A_\mathrm{FB}$ assuming 50 signal events.}
  \label{fig:RateComparison}
\end{figure}

For both the SHM and stream, the asymmetry is significantly underestimated when using only the simple forward-backward ($N=2$) discretisation. This matches the analysis of the previous sections, which demonstrated that the discretised velocity distribution tends to lead to a leakage of events from the forward to the backward direction. However, for both distributions, the asymmetry obtained from the discretised distribution rapidly approaches the true value with increasing $N$. For the stream distribution, the approximation converges more rapidly than for the SHM. This is because of the strong directionality of the stream velocity distribution, so only relatively few bins are required to capture a simple forward-backward asymmetry. Even with only 4 angular bins, the error in $A_\mathrm{FB}$ for both velocity distributions is less than 10\% and lies within the $1\sigma$ statistical uncertainty expected from a sample of 50 signal events. 

\section{Discussion}
\label{sec:discussion}

We have demonstrated that for smooth SHM-like distributions, the discretised velocity distribution allows us to obtain a good approximation to the integrated Radon transform (IRT) with relatively few angular bins. Though the $N=2$ discretisation is too simple an approximation, for 3 or more bins the discretised distribution leads to a number of events in each bin which matches the true number within statistical uncertainties (assuming 50 signal events). This means that we should be able to parametrise the speed distributions in each angular bin $f^k(v)$ and account for astrophysical uncertainties without introducing a large error in the number of observed events. For comparison, we have also considered a more extreme and highly-directional stream distribution. The disagreement between the true recoil spectrum and that obtained from the discretised velocity distribution is significantly larger in this case. Though increasing the number of bins $N$ improves the agreement, a rather large number of bins ($N \sim 18$) may be needed to accurately describe the stream distribution. 

The results we have presented here have assumed an idealised future directional detector. However, application of these results is not necessarily as straightforward in a realistic experiment. Conversion from $\vmin$ to $E_R$ requires knowledge of the DM mass $m_\chi$ and attempting to fit both the velocity distribution and mass simultaneously can lead to spurious results \cite{Peter:2011, Kavanagh:2012}. Furthermore, for comparison with experimental data, we must take into account the fact that experiments have finite angular resolution, typically in the range 20$\,^{\circ}$-80$\,^{\circ}$, with higher resolution at high energy \cite{Billard:2012}. Finally, throughout this work, we have assumed that the basis for discretising the velocity distribution is aligned with the peak of the underlying velocity distribution (i.e. $\mathbf{v}_e$ aligned along $\theta' = 0$). However, this is not known \textit{a priori} and a reliable method of selecting the orientation of the angular basis is required.

In spite of these open issues, the work presented here is general and conservative. We have demonstrated that the stream distribution is more poorly fit than the SHM distribution with few angular bins. However, the stream is an extreme example, as discussed in Sec.~\ref{sec:compare}, and it is unlikely the discretisation error will be any greater than observed in that case. Other possibilities for the velocity distribution would tend to be less directional than the stream example, leading to better agreement between the exact and approximate rates. For a dark disk, the low value of $v_e$ means that the velocity distribution appears almost isotropic in the lab frame, which will only reduce the angular discretisation error. A misalignment between $\mathbf{v}_e$ and $\theta' = 0^\circ$ would also lead to a distribution more isotropic in $\theta'$. Similarly, including finite angular resolution will smear the velocity distribution (in a style similar to the discretisation), reducing the directional asymmetry and therefore the discretisation error. We have also demonstrated that this method is applicable in detectors where sense recognition is not possible.

For a realistic analysis, all of these issues should be taken into account. Based on mock (or future) data, the optimal direction for $\theta' = 0^\circ$ should be selected (potentially based on the observed median recoil direction \cite{Morgan:2005,Green:2010b}). An appropriate number of bins $N$ should then be chosen, which may be influenced by the angular resolution of the experiment and the number of observed events. In some scenarios, such as for strongly peaked signals, deviations from the smooth SHM may be obvious from the data. In this case, a definition of the angular bins different from that used here (including differently-sized bins) may be optimal to reduce the discretisation error. Using the results of Appendix~\ref{app:RadonDeriv}, such a scenario can be straightforwardly accommodated, although we leave the issue of choosing the number and size of the bins to future studies.

In order to make more quantitative statements, we have focused on a specific scenario, in which 50 signal events (and 1 background event) are observed. This has allowed us to compare the typical Poisson uncertainties on the number of observed events with the error induced by using the discretised velocity distribution. Though a signal consisting of 50 events is somewhat arbitrary, it is representative of the lower limit on the number of events required before such an analysis would be reasonable. As the number of signal events increases, the typical statistical uncertainty decreases. Eventually, this uncertainty becomes smaller than the error between the approximate and exact results. Increasing the number of angular bins would then reduce the discretisation error and increase the statistical uncertainty on the number of events in each bin, reducing any bias which may be induced by the discretisation. This therefore provides a natural scheme for deciding how many angular bins are required, based on the number of signal events observed.

Once the number of bins $N$ has been fixed and a suitable parametrisation for the radial functions $f^k(v)$ has been chosen, the full parameter space (of both DM particle physics parameters and astrophysics parameters) should be fit to the data. We note in particular that by summing the radial functions $f^k(v)$, the 1-dimensional speed distribution is recovered, meaning that non-directional experiments can easily be included in this framework. As emphasised in this section, the fitting process will be highly dependent on the underlying DM and experimental parameters, as well as on the chosen parametrisation for $f^k(v)$. We have therefore focused here on the discretisation of $f(\mathbf{v})$ and left more involved investigations for future work.

\section{Conclusions}
\label{sec:conclusions}

In this work, we have presented an angular basis which can be used to parametrise the DM velocity distribution. This involves discretising the velocity distribution into $N$ angular bins:
 \begin{equation}
f(\textbf{v}) = f(v, \cos\theta', \phi') =
\begin{cases}
f^1(v) & \textrm{ for } \theta' \in \left[ 0, \pi/N\right]\,, \\
f^2(v) & \textrm{ for } \theta' \in \left[ \pi/N, 2\pi/N\right]\,, \\
 & \vdots\\
f^k(v) & \textrm{ for } \theta' \in \left[ (k-1)\pi/N, k\pi/N\right]\,, \\
 & \vdots\\
f^N(v) & \textrm{ for } \theta' \in \left[ (N-1)\pi/N, \pi\right]\,. \\
\end{cases}
\end{equation} 
In Appendices~\ref{app:RadonDeriv} and \ref{app:Radon}, we have provided recipes for calculating the corresponding directionally-integrated Radon transforms (IRTs), which appear in the directional event rate. Alternative methods for parametrising the DM velocity distribution (such as Refs.~\cite{Alves:2012, Lee:2014}, and in particular those based on spherical harmonics) may lead to negative values $f(\mathbf{v})$, leading to unphysical distribution functions. It is not clear what affect this might have on parameter reconstruction or how this problem can be mitigated. The advantage of the basis presented here is that it guarantees that the resulting distribution function is everywhere positive and therefore physical.

We have investigated the possible size of discretisation errors in the directional event rate by comparing the IRT in each bin obtained from the discretised and full velocity distributions. The simplest possible discretisation ($N=2$) is unsuitable for fitting to data, as it substantially underestimates the forward scattering rate, while underestimating the backwards scattering rate. However, starting from $N=3$, the standardly-assumed SHM can be well fit by the discretised velocity distribution. We have shown that once the DM nature of the signal is confirmed (with around 50 events), the $N=3$ discretisation allows us to calculate the number of events in each angular bin and obtain good agreement with the true event numbers (within statistical uncertainties). Increasing the number of bins to $N=5$ improves this agreement, giving discretisation errors in the range 10-50\% depending which angular bin is considered. 

For comparison, we also consider an extreme, highly directional stream distribution. In this case, the discretisation error is much larger and exceeds the statistical uncertainties in the $N=5$ case. An estimated $N=18$ angular bins would be required to accurately map the full velocity distribution. However, increasing $N$ does reduce the discrepancy between the exact and approximate results. In addition, with as few as $N=5$ bins, it is possible to distinguish between the SHM and stream distribution. This suggests that even though the discretised distribution gives a poor approximation to the stream, it may be useful in distinguishing different distributions, even for relatively few bins.

Finally, we have considered the forward-backward asymmetry $A_{\mathrm{FB}}$ in the event rate, calculated using both the full velocity distribution and the discretised velocity distribution. The latter calculation rapidly converges to the correct value with increasing $N$. For both the SHM and stream distributions, the error in $A_\mathrm{FB}$ is below $10\%$ for $N \geq 4$ bins, within the statistical error expected for 50 signal events. We have also demonstrated that the method presented here can still be used when head-tail discrimination of the recoil tracks is not possible, in which case $N \geq 3$ bins are required.

In this work, we have only considered discretising the angular component of the DM velocity distribution, leaving the $N$ radial functions $f^{k}(v)$ fixed to their `true' values. In future work, it will be necessary to combine the discretisation presented here with a parametrisation of these radial functions (such as the parametrisations of Refs.~\cite{Peter:2011,Kavanagh:2012, Kavanagh:2014}), in order to reconstruct the velocity distribution (and other DM parameters) from mock data. In addition, several questions are still to be addressed, including how to decide on the optimal `forward' direction for the discretisation, and how realistic angular resolution would impact the results presented here. However, these additional considerations are only expected to improve the agreement between the true and approximate event rates. This work represents an initial step towards a parametrisation of the full DM velocity distribution, which would allow future data from directional experiments to be analysed without astrophysical uncertainties and which would potentially allow the velocity distribution itself to be probed and reconstructed.

\acknowledgments

The author thanks Anne M. Green, Julien Billard and Samuel K. Lee for helpful discussions on the subject of directional detection. The author also thanks Anne M. Green and Ciaran A. J. O'Hare for critical comments on this manuscript. The author is supported by the European Research Council (ERC) under the EU Seventh Framework Programme (FP7/2007- 2013)/Erc Starting Grant (agreement n. 278234 -- 'NewDark' project).

\begin{appendix}

\section{Calculating the Radon transform}
\label{app:RadonDeriv}
In this appendix, we derive the Radon transform corresponding to the discretised velocity distribution. The final result can be found in Appendix~\ref{app:Radon}. The Radon transform of $f(\textbf{v})$ is defined by:

\begin{equation}
\hat{f}(\vmin, \qhat) = \int_{\mathbb{R}^3} f(\mathbf{v}) \, \delta\left(\mathbf{v}\cdot\qhat - \vmin\right) \, \mathrm{d}^3\mathbf{v}\,.
\end{equation}
We write the coordinates of the velocity and recoil momentum as

\begin{align}
\begin{split}
\mathbf{v} &= v\left(\sin\theta'\cos\phi', \sin\theta'\sin\phi', \cos\theta'\right) \\
\qhat &= \left(\sin\theta\cos\phi, \sin\theta\sin\phi, \cos\theta\right) \,,
\end{split}
\end{align}
and consider here only the azimuthally integrated Radon transform. We write this as $\hat{f}(\vmin, \cos\theta)$, which is given by

\begin{align}
\begin{split}
\hat{f}(\vmin, \cos\theta)  &= \int_{0}^{2\pi} \hat{f}(\vmin, \cos\theta,\phi) \, \mathrm{d}\phi \\
&=  \int_{0}^{2\pi} \left( \int_{\mathbb{R}^3} f(\mathbf{v}) \, \delta\left(\mathbf{v}\cdot\qhat - \vmin\right) \, \mathrm{d}^3\mathbf{v}\right) \, \mathrm{d}\phi \\
& = \int_{\mathbb{R}^3} f(\mathbf{v}) \left(\int_{0}^{2\pi}   \, \delta\left(\mathbf{v}\cdot\qhat - \vmin\right) \, \mathrm{d}\phi \right) \, \mathrm{d}^3\mathbf{v}\\
&\equiv \int_{\mathbb{R}^3} f(\mathbf{v}) \, D\left(\vmin, \cos\theta, \mathbf{v} \right) \, \mathrm{d}^3\mathbf{v}\,.
\end{split}
\end{align}
We expand the $\delta-$function explicitly in terms of the angular coordinates:

\begin{equation}
\delta\left(\mathbf{v}\cdot\qhat - \vmin\right) = \frac{1}{v}\delta\left(\sin\theta\sin\theta'\cos(\phi - \phi') + \cos\theta\cos\theta' - \vmin/v  \right) \equiv \frac{1}{v}\delta\left(g(\phi)\right)\,.
\end{equation}
We rewrite the argument of the $\delta-$function as a function $\phi$:
\begin{equation}
\label{eq:deltadecomp}
\delta\left(  g(\phi) \right) = \sum_{i} \frac{\delta(\phi - \phi_i)}{\left| g'(\phi_i) \right|}\,.
\end{equation}
Here, we sum over those values of $\phi_i$ satisfying $g(\phi_i) = 0$:

\begin{equation}
\label{eq:gamma}
\cos(\phi_i - \phi') = \frac{\beta - \cos\theta\cos\theta'}{\sin\theta\sin\theta'} \equiv \alpha\,,
\end{equation}
where we have also defined $\beta = \vmin/v$. The solutions for $\phi \in [0, 2\pi]$ are:

\begin{align}
\phi_1 &= \phi' + \acos\alpha, & \textrm{ for } &  \phi' \in \left[0, 2\pi - \acos\alpha\right] \notag\\
\phi_2 &= \phi' + 2\pi - \acos\alpha, & \textrm{ for } & \phi' \in \left[0, \acos\alpha\right] \notag\\
\phi_3 &= \phi' + \acos\alpha -2\pi, & \textrm{ for } & \phi' \in \left[2\pi - \acos\alpha, 2\pi\right] \notag\\
\phi_4 &= \phi' -\acos\alpha, & \textrm{ for } & \phi' \in \left[\acos\alpha, 2\pi\right] \,.
\end{align}
We note that these solutions exist only for $\beta \in \left[0,1\right]$ (or equivalently $v > \vmin$) and for $\alpha \in [-1,1]$, otherwise Eq.~\ref{eq:gamma} cannot be satisfied. If these constraints are satisfied, there exist exactly 2 solutions for a given value of $\phi'$ and therefore 2 $\delta$-functions in Eq.~\ref{eq:deltadecomp}.

For the derivative of $g(\phi)$ we obtain
\begin{align}
g'(\phi) = -\sin\theta\sin\theta'\sin(\phi-\phi')\,.
\end{align}
Substituting the values of $\phi_{1,2,3,4}$, we see that
\begin{align}
|g'(\phi_{1,2,3,4})| = \sqrt{\left(\sin\theta\sin\theta'\right)^2 - \left(\beta - \cos\theta\cos\theta'\right)^2}\,.
\end{align}
Each of the two $\delta$-functions therefore contributes the same amount to the integral, regardless of the value of $\phi'$.  We can now perform the integral over $\phi$:

\begin{align}
\begin{split}
D\left(\vmin, \cos\theta, \mathbf{v} \right)  &= \frac{1}{v} \int_{0}^{2\pi}   \, \delta\left(g(\phi)\right) \, \mathrm{d}\phi \\
&= \frac{2 C(\alpha)}{v\sqrt{\left(\sin\theta\sin\theta'\right)^2 - \left(\beta - \cos\theta\cos\theta'\right)^2}}\Theta(v - \vmin) \\
&\equiv \frac{2}{v}C(\alpha) I(\beta, \cos\theta, \cos\theta') \Theta(v - \vmin) 
\end{split}
\end{align}
where $C(\alpha) = 1$ for $\alpha \in [-1,1]$ and vanishes otherwise. We note that $D$ is independent of the azimuthal angle $\phi'$. The constraint $\alpha \in [-1,1]$ is satisfied for $\cos\theta' \in [x_-, x_+]$, where

\begin{equation}
x_\pm= \beta \cos\theta \pm \sqrt{1-\beta^2}\sin\theta \,.
\end{equation}
We therefore obtain
\begin{equation}
\hat{f}(\vmin, \cos\theta) = 2 \int_{x_-}^{x_+} \int_{\vmin}^{\infty}  f(v, \cos\theta') I(\beta, \cos\theta, \cos\theta')\, v \,\mathrm{d}v\, \mathrm{d}\cos\theta'\,,
\end{equation}
where we have performed the $\phi'$ integral over the velocity distribution, and defined
\begin{equation}
f(v, \cos\theta') = \int_{0}^{2\pi} f(v, \cos\theta', \phi') \, \mathrm{d}\phi'\,,
\end{equation}
because $I$ does not depend on $\phi'$.  We note from this that the azimuthally-integrated Radon transform is unaffected by the exact $\phi'$-dependence of $f(\mathbf{v})$. Instead, the azimuthally-integrated Radon transform depends only on the azimuthally-integrated velocity distribution. For the framework considered here, then, we can assume that $f(\mathbf{v})$ is independent of $\phi'$ without loss of generality.

We will consider a velocity distribution which is discretised into $N$ angular pieces: 
\begin{equation}
f(\textbf{v}) = f(v, \cos\theta', \phi') =
\begin{cases}
f^1(v) & \textrm{ for } \theta' \in \left[ 0, \pi/N\right]\,, \\
f^2(v) & \textrm{ for } \theta' \in \left[ \pi/N, 2\pi/N\right]\,, \\
 & \vdots\\
f^k(v) & \textrm{ for } \theta' \in \left[ (k-1)\pi/N, k\pi/N\right]\,, \\
 & \vdots\\
f^N(v) & \textrm{ for } \theta' \in \left[ (N-1)\pi/N, \pi\right]\,. \\
\end{cases}
\end{equation}
We would then ultimately like to calculate the directionally-integrated Radon transform $\hat{f}^j$, integrated over the same bins in $\cos\theta$ as for the velocity distribution,

\begin{align}
\label{eq:fj1}
\begin{split}
\hat{f}^j(\vmin) &=  \int_{\cos(j\pi/N)}^{\cos((j-1)\pi/N)} \hat{f}(\vmin, \cos\theta) \, \mathrm{d}\cos\theta \\
&= 2 \int_{a_{j}}^{a_{j-1}} \int_{\vmin}^{\infty}  \left( \int_{x_-}^{x_+} f(v,\cos\theta') I(\beta, \cos\theta, \cos\theta')  \,\mathrm{d}\cos\theta' \right) \, v \,\mathrm{d}v \,\mathrm{d}\cos\theta \,,
\end{split}
\end{align}
Here, we have introduced the notation $a_k = \cos(k\pi/N)$ and defined $f(v, \cos\theta') = 2\pi f(v, \cos\theta', \phi')$. We divide the integration range for $\cos\theta'$ into different regions:

\begin{align}
\begin{split}
\cos\theta' &\in [x_-, a_{k_--1}] \\
\cos\theta' &\in [a_{k_--1}, a_{k_--2}]\\
\vdots&\\
\cos\theta' &\in [a_{k_+ + 1}, a_{k_+}]\\
\cos\theta' &\in [a_{k_+}, x_+]\,,
\end{split}
\end{align}
where $k_\pm$ are defined such that:

\begin{equation}
x_\pm \in [a_{k_\pm}, a_{k_\pm - 1}]\,. 
\end{equation}
These regions are chosen such that $f(v, \cos\theta')$ is independent of $\cos\theta'$ within each region. We note that if $k_+ = k_-$, there is only one region: $\cos\theta' \in [x_-, x_+]$. Using these definitions we can therefore rewrite the term in brackets in Eq.~\ref{eq:fj1} as

\begin{align}
\begin{split}
\label{eq:stepone}
&\int_{x_-}^{x_+} f(v,\cos\theta') I(\beta, \cos\theta, \cos\theta')  \,\mathrm{d}\cos\theta' \\
&=2\pi f^{k_-}(v) \int_{x_-}^{a_{k_--1}}  I(\beta, \cos\theta, \cos\theta')  \,\mathrm{d}\cos\theta' \\
&+\sum_{k=k_- - 1}^{k_+ +1} 2 \pi f^{k}(v)  \int_{a_k}^{a_{k-1}}  I(\beta, \cos\theta, \cos\theta')  \,\mathrm{d}\cos\theta'  \\
&+ 2\pi f^{k_+}(v) \int_{a_{k_+}}^{x_+}  I(\beta, \cos\theta, \cos\theta')  \,\mathrm{d}\cos\theta'\,.
\end{split}
\end{align}
We can now explicitly perform the integral over $\cos\theta'$,

\begin{align}
\begin{split}
&\int_{a_k}^{a_{k-1}} I(\beta, \cos\theta, \cos\theta')  \,\mathrm{d}\cos\theta' \\
&= \int_{a_{k}}^{a_{k-1}} \frac{\mathrm{d}\cos\theta'}{\sqrt{(\sin\theta\sin\theta')^2 - (\beta - \cos\theta\cos\theta')^2}} \\
&= \left[ \asin\left(  \frac{\cos\theta' - \beta\cos\theta}{\sqrt{1-\beta^2}\sin\theta} \right)    \right]_{a_k}^{a_{k-1}} \\
&=  F(a_{k-1}, \cos\theta) - F(a_k,\cos\theta)\,,
\end{split}
\end{align}
where we have defined
\begin{equation}
F(y, \cos\theta) =  \asin\left(  \frac{y - \beta\cos\theta}{\sqrt{1-\beta^2}\sin\theta} \right) \,.
\end{equation}
We note that
\begin{equation}
F(x_\pm, \cos\theta) = \pm \frac{\pi}{2}\,,
\end{equation}
for all $\cos\theta$.
We also note that $F(y, \cos\theta)$ can be integrated analytically:

\begin{align}
\label{eq:F}
\begin{split}
&\int^x F(y, \cos\theta) \, \cos\theta\\
&= x \asin\left(\frac{y-\beta x}{\sqrt{1-x^2}\sqrt{1-\beta^2}}\right)\\
&\, + y \atan\left(\frac{x-y\beta}{\sqrt{t}}\right) \\
&\, + \frac{1}{2} \atan\left(\frac{1 - y^2 - \beta^2 - x + y\beta(1+x)}{(y-\beta) \sqrt{t}}\right) \\
&\, - \frac{1}{2} \atan\left(\frac{1 - y^2 - \beta^2 + x - y\beta(1-x)}{(y+\beta) \sqrt{t}}\right)\\
&\equiv J(x,y) + C
\end{split}
\end{align}

where it is understood that $J(x,y)$ is a function of $\beta$ and where
\begin{equation}
t = (1-x^2)(1-\beta^2) - (y - \beta x)^2\,.
\end{equation}

It is now possible to also complete the integration over $\cos\theta$. Assuming that $k_\pm$ do not change over a range $\cos\theta \in [b_i, b_{i+1}]$, we can combine Eqs.~\ref{eq:stepone} and \ref{eq:F} to obtain

\begin{align}
\label{eq:steptwo}
\begin{split}
&\int_{b_i}^{b_{i+1}} \left(\int_{x_-}^{x_+} f(v, \cos\theta') I(\beta, \cos\theta, \cos\theta') \, \mathrm{d}\cos\theta'  \right)\, \mathrm{d}\cos\theta \\
&= 2\pi f^{k_-}(v) \left(  \frac{\pi}{2}(b_{i+1} - b_i) + J(b_{i+1}, a_{k_- - 1})  - J(b_{i}, a_{k_-- 1}) \right) \\
&+\sum_{k=k_- - 1}^{k_+ +1} 2 \pi f^{k}(v) \left( J(b_{i+1},a_{k-1})  - J(b_{i},a_{k-1}) - J(b_{i+1},a_{k}) + J(b_{i},a_{k})\right) \\
&+ 2\pi f^{k_+}(v) \left( \frac{\pi}{2}(b_{i+1} - b_i) - J(b_{i+1}, a_{k_+})  - J(b_{i}, a_{k_+}) \right)\,.
\end{split}
\end{align}
However, we must take into account the fact that the values of $x_\pm$ and therefore of $k_\pm$ depend on $\beta$ and $\cos\theta$. We now discuss how to divide up the integration regions of $v$ and $\cos\theta$ such that we can apply Eq.~\ref{eq:steptwo}.

\begin{figure}[t]
  \centering
  \includegraphics[width=0.5\textwidth]{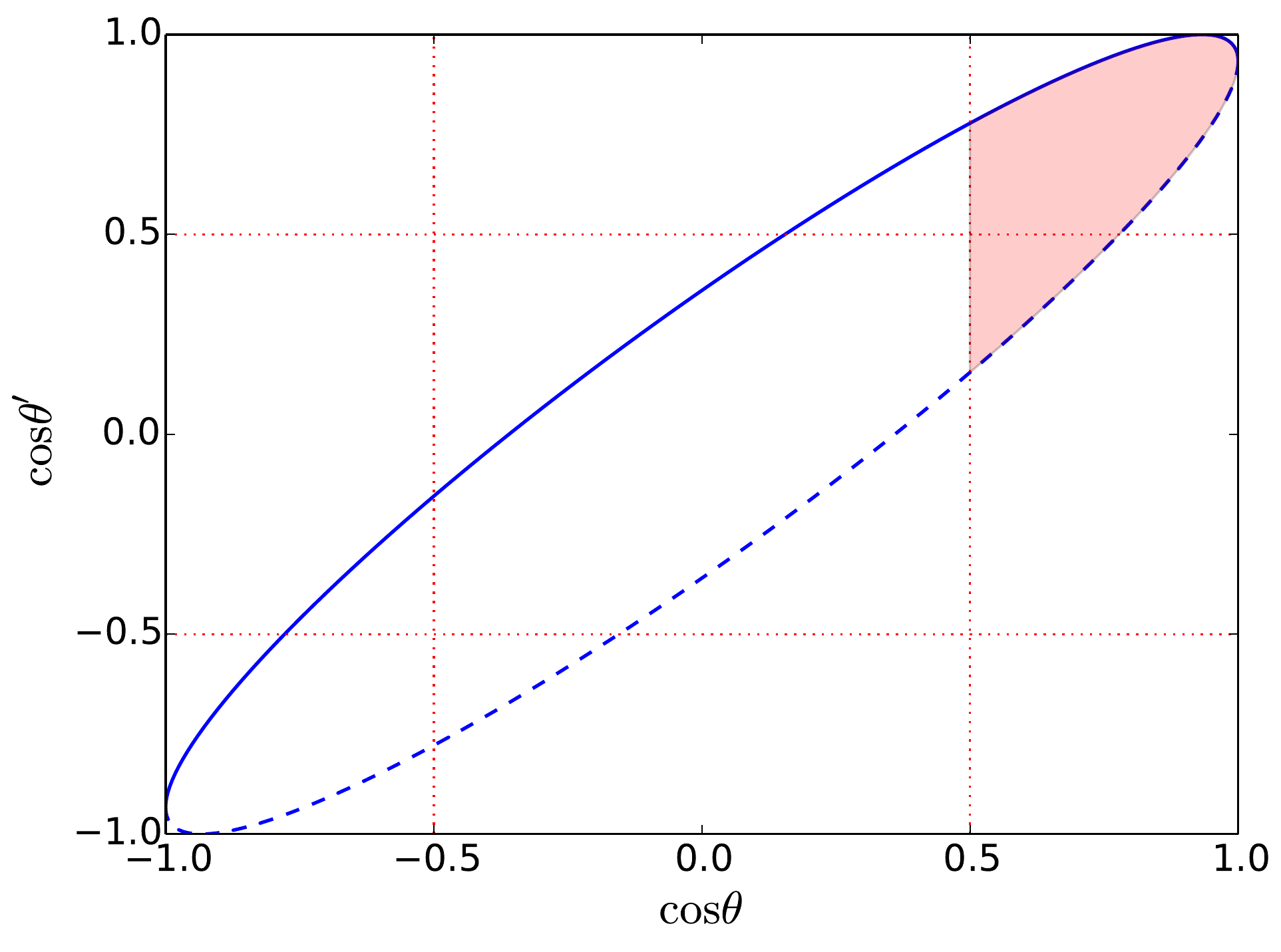}
\caption{Integration limits for Eq.~\ref{eq:fj1} as a function of $\cos\theta$, for a fixed value of $\beta$. The limits $x_+$ and $x_-$ are shown as solid and dashed lines respectively. The dotted vertical and horizontal lines mark the values $\cos(n \pi/N)$ for $n = 0, 1, 2, 3$ and $N=3$. If we wish to calculate $\hat{f}^1(\vmin)$, we must perform the angular integral over the shaded region.}
\label{fig:integlimits1}
\end{figure}

As an illustration, we show in Fig.~\ref{fig:integlimits1} the values of $x_+ (x_-)$ as solid (dashed) lines as a function of $\cos\theta$, for a fixed value of $\beta$. In evaluating Eq.~\ref{eq:fj1}, we wish to integrate over the region enclosed by the solid and dashed lines, for the relevant range of $\cos\theta$.  As an example, we show with horizontal and vertical dotted lines the edges of the discretised regions in $\cos\theta$ and $\cos\theta'$ for $N=3$. That is, the dotted lines show the values $\cos(n \pi/N)$ for $n = 0, 1, 2, 3$. If we wish to evaluate $\hat{f}^1(\vmin)$, we need to integrate over the shaded region in Fig.~\ref{fig:integlimits1}.

\begin{figure}[t]
  \centering
  \includegraphics[width=0.5\textwidth]{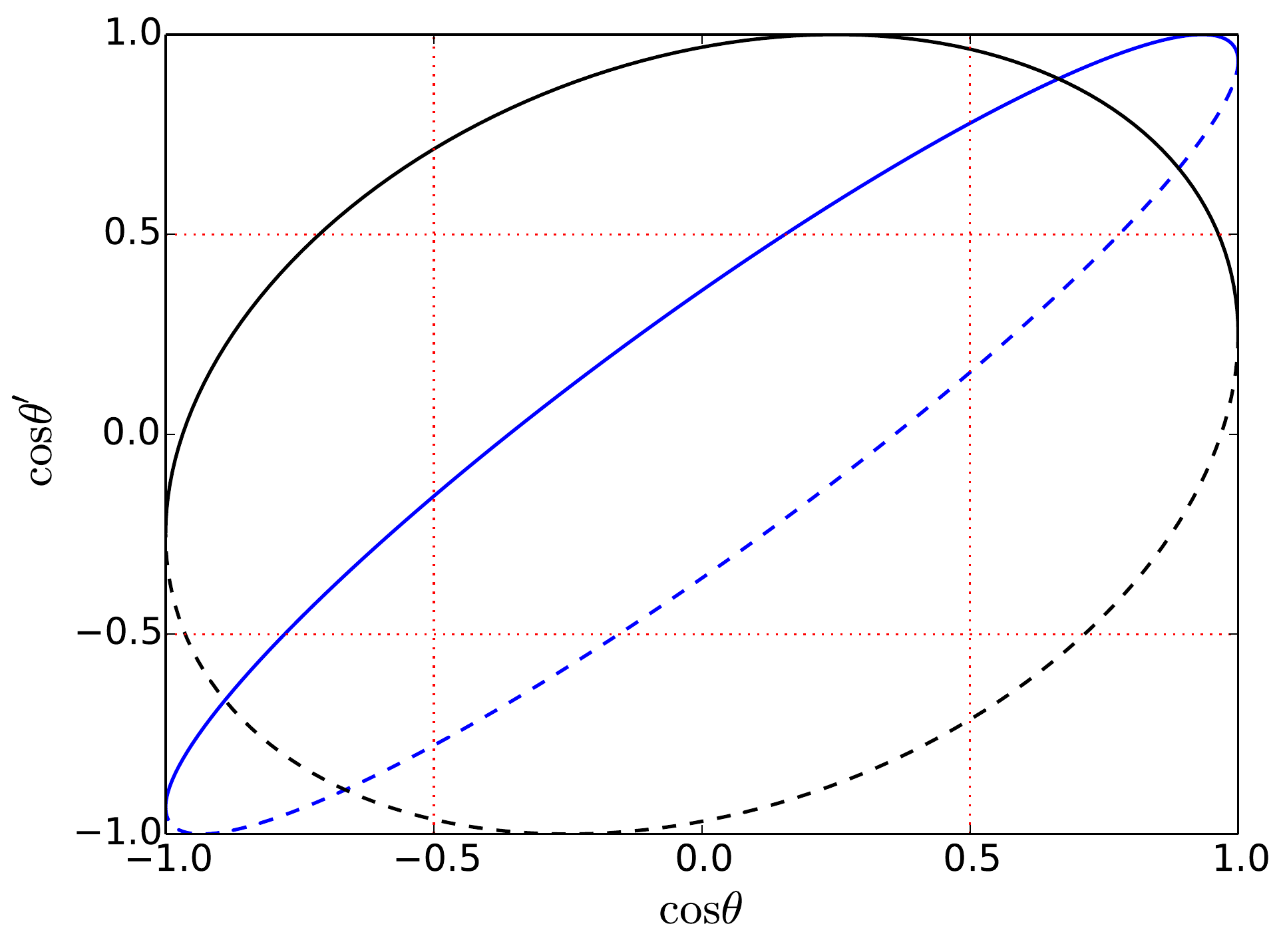}
\caption{As Fig.~\ref{fig:integlimits1}, but for two different values of $\beta$. The black lines show $x_\pm$ for a value of $\beta$ in the range $\beta < 1/2$ while blue lines show $x_\pm$ for a value in the range $\beta > \sqrt{3}/2$.}
  \label{fig:integlimits2}
\end{figure}

In Fig.~\ref{fig:integlimits2}, we show $x_\pm$ for different values of $\beta$. We give two examples, for two different ranges of $\beta$: $\beta < 1/2$ (black) and $\beta > \sqrt{3}/2$ (blue). The values of $k_\pm$ (and therefore the relevant $f^k(v)$) will clearly depend on the value of $\beta$, so we must be careful to account for this properly.

Calculation of the values of $k_\pm$ and the corresponding bin edges in $\cos\theta$ ($\left\{b_i\right\}$) is involved, though not technically difficult. We therefore sketch the procedure here. We begin with the definition of $x_+$:

\begin{equation}
\label{eq:xplusdef}
x_+ = x_+(\cos\theta) = \beta \cos\theta + \sqrt{1-\beta^2}\sin\theta\,.
\end{equation}
By straight-forward differentiation, we observe that this function has a maximum at $\cos\theta = \beta$. For concreteness, we consider $\beta$ in the range,

\begin{equation}
\beta \in \left[\cos(n\pi/N), \cos((n-1) \pi/N) \right] \qquad \textrm{ for } n = 1, 2, ..., N/2 \,.
\end{equation}
We must consider three separate cases, depending on $j$ (i.e. which $\cos\theta$ bin we are considering):

\begin{enumerate}[(i)]

\item $j \leq (n-1)$

In this case, $\cos\theta > \beta$, meaning that $\partial x_+ / \partial \cos\theta < 0$, so $x_+$ is monotonically decreasing in $\cos\theta$.

\item $j \geq (n+1)$

In this case, $\cos\theta < \beta$, meaning that $\partial x_+ / \partial \cos\theta > 0$, so $x_+$ is monotonically increasing in $\cos\theta$.

\item $j =  n$

In this case, the maximum of $x_+$ lies in the $j^\mathrm{th}$ bin.

\end{enumerate}

For each value of $j$, we can now determine the maximum and minimum value of $x_+$ within that $\cos\theta$ bin by using the definition in Eq.~\ref{eq:xplusdef} . This allows us to determine the value of $k_+$ (i.e. the $\cos\theta'$ bin into which $x_+$ falls). In fact, it can be shown that for each value of $j$ (apart from $j=n$), $x_+$ crosses one of the $\cos\theta'$ bin edges exactly once, when $\cos\theta = \cos\theta_+^j$, satisfying

\begin{equation}
x_+(\cos\theta_+^j) = \cos(l \pi/N)\,,
\end{equation}
for some $l = 1,2,...,N$. This means that for a given $j$, there are two distinct regions in $\cos\theta$ ($\cos\theta < \cos\theta_+^j$ and $\cos\theta > \cos\theta_+^j$) each with a different value of $k_+$. 

We can perform a similar analysis for $x_-$ to obtain the values of $k_-$ for a given $n$ and $j$, as well as the values of $\cos\theta_-^j$, where the curve $x_-$ crosses from one bin to the next. However, we can only apply Eq.~\ref{eq:steptwo} if both $k_+$ and $k_-$ do not change over the range $\cos\theta \in [b_i, b_{i+1}]$. It is clear then that we must subdivide each $\cos\theta$ bin into 3 smaller bins. The edges of these sub-bins will be $\cos\theta_+^j$ and  $\cos\theta_-^j$, at which the value of either $k_+$ or $k_-$ changes. 

Finally, we need to determine which of these crossings $\cos\theta_\pm^j$ occurs first. The crossings occur at the same value of $\cos\theta$ if $\cos\theta_+^j$ = $\cos\theta_-^j$, which occurs for $\beta = \cos\left((n-1/2)\pi/N\right)$. For each value of $n$, we therefore need to distinguish two regimes, depending on which of the crossings $\cos\theta_\pm^j$ is larger. It is helpful then to further subdivide the range of $\beta$, writing
\begin{equation}
\beta \in \left[\cos(m\pi/(2N)), \cos((m-1) \pi/(2N)) \right] \qquad \textrm{ for } m = 1, 2, ..., N \,,
\end{equation}
from which we obtain 
\begin{equation}
n = \begin{cases}
(m+1)/2 & \textrm{ for } m \textrm{ odd}\\
m/2  & \textrm{ for } m \textrm{ even}\,.
\end{cases}
\end{equation} 
We can now determine which of $\cos\theta_\pm^j$ is greater, depending on whether $m$ is even or odd. For a given value of $m$ then, the values of the sub-bin edges in $\cos\theta$ ($\left\{b_i\right\}$)  are now well determined, along with the values of $k_\pm$ within each sub-bin. This allows us to apply Eq.~\ref{eq:steptwo} to each of these sub-bins.

Finally, we can perform the integral over $v$ described in Eq.~\ref{eq:fj1}. However, we note that we should sub-divide the range of integration, depending on the value of $m$. That is, the integral over $v$ becomes,

\begin{equation}
\int_{\vmin}^{\infty} \, \mathrm{d}v \rightarrow \sum_{m = 1}^N \int_{\vmin/a_{(m-1)/2}}^{\vmin/a_{m/2}} \, \mathrm{d} v\,,
\end{equation}
where we remind the reader that $a_k = \cos(k\pi/N)$. For each term in this sum, the value of $m$ does not change within the range of integration, so we can apply the results which have been obtained so far. We give the full expression for the IRT, as well as the values of $k_\pm$ and $\cos\theta_\pm^j$ (calculated as described here) in Appendix~\ref{app:Radon}.

\section{The directionally-integrated Radon transform (IRT)}
\label{app:Radon}

The discretised velocity distribution, with $N$ angular bins, is defined as

\begin{equation}
f(\textbf{v}) = f(v, \cos\theta', \phi') =
\begin{cases}
f^1(v) & \textrm{ for } \theta' \in \left[ 0, \pi/N\right]\,, \\
f^2(v) & \textrm{ for } \theta' \in \left[ \pi/N, 2\pi/N\right]\,, \\
 & \vdots\\
f^k(v) & \textrm{ for } \theta' \in \left[ (k-1)\pi/N, k\pi/N\right]\,, \\
 & \vdots\\
f^N(v) & \textrm{ for } \theta' \in \left[ (N-1)\pi/N, \pi\right]\,. \\
\end{cases}
\end{equation}
The integrated Radon transform (IRT), integrated over the same angular bins, is defined as

\begin{equation}
\hat{f}^j(v_\textrm{min}) = \int_{\phi = 0}^{2\pi} \int_{a_{j}}^{a_{j-1}} \hat{f}(v_\textrm{min}, \hat{\textbf{q}})\, \mathrm{d}\cos\theta \,\mathrm{d}\phi\,,
\end{equation}
where have defined the shorthand $a_k = \cos(k\pi/N)$.
The IRT is given in terms of the distribution functions $f^k(v)$ as

\begin{align}
\begin{split}
\hat{f}^j(\vmin) &= 4\pi\sum_{m = 1}^N  \int_{\vmin/a_{(m-1)/2}}^{\vmin/a_{m/2}} v \, \textrm{d}v\sum_{i = 0,1,2} \Big[\frac{\pi}{2} (f^{k_+^{i}}(v) + f^{k_-^{i}}(v)) \left(b_{i+1} - b_i\right)   \\
& + (1-\delta_{k_+^{i} k_-^{i}}) f^{k_-^{i}}(v)\left( J(b_{i+1}, a_{k_-^{i} - 1}) -   J(b_{i}, a_{k_-^{i} - 1})  \right) \\
& -(1-\delta_{k_+^{i} k_-^{i}}) f^{k_+^{i}}(v)\left( J(b_{i+1}, a_{k_+^{i}}) -   J(b_{i}, a_{k_+^{i}})  \right) \\
& + \sum_{k = k_+^{i} + 1}^{k_-^{i}- 1} f^{k}(v)\left(J(b_{i+1},a_{k-1})  - J(b_{i},a_{k-1}) \right. \\
&\left. \qquad \qquad \qquad- J(b_{i+1},a_{k}) + J(b_{i},a_{k})\right) \big]\, .
\end{split}
\end{align}
The function $J(x,y)$ depends implicitly on the speed $v$ and is defined below. For each value of $j$, we integrate over N ranges in $v$, labeled by the index $m$. We also sum over three `bins', labeled by the index $i = 0,1,2$, which are distinguished by their values of the integers $k_\pm^{i}$. We note that the values of $k_\pm^{i}$ will depend on the values of $j$ and $m$ under consideration, as well as the index $i$. We write $\mathbf{k_{\pm}} = (k_\pm^{0}, k_\pm^{1}, k_\pm^{2})$ and give the values below.

For odd $m$, we define $n = (m+1)/2$ and bin edges

\begin{align}
\begin{split}
b_0 &= \cos\left(j \pi/N\right)\\
b_1 &= \cos\left(\gamma + (j + n - 1)\pi/N\right) \\
b_2 &= \cos\left(\gamma + (j -n)\pi/N\right) \\
b_3 &= \cos\left((j-1) \pi/N\right)\,,
\end{split}
\end{align}
with $\gamma = \acos\left(\vmin/v\right)$. The corresponding $\mathbf{k}_\pm$ values are:

\begin{equation}
\mathbf{k}_+ = 
\begin{cases}
\left(n - j, n - j, n - j + 1 \right) & \textrm{ for } j \leq (n-1) \\
\left(1,1,1\right) & \textrm{ for } j = n \\
\left(j-n+1, j-n+1, j-n \right) & \textrm{ for } j \geq (n+1)\,, \\
\end{cases}
\end{equation}
and
\begin{equation}
\mathbf{k}_- = 
\begin{cases}
\left(n+j, n+j -1, n+j-1 \right) & \textrm{ for } j \leq (N-n) \\
\left(N,N,N\right) & \textrm{ for } j = (N+1-n)\\
\left(2N - n -j +1, 2N - n - j +2, 2N-n-j + 2 \right) & \textrm{ for } j \geq (N+2-n)\,. \\
\end{cases}
\end{equation}

For even $m$, we write $n= m/2$ and bin edges
\begin{align}
\begin{split}
b_0 &= \cos\left(j \pi/N\right)\\
b_1 &= \cos\left(\gamma + (j - n)\pi/N\right) \\
b_2 &= \cos\left(\gamma + (j+n -1)\pi/N\right) \\
b_3 &= \cos\left((j-1) \pi/N\right)\,.
\end{split}
\end{align}
The corresponding $\mathbf{k}_\pm$ values are:
\begin{equation}
\mathbf{k}_+ = 
\begin{cases}
\left(n - j, n - j+1, n - j + 1 \right) & \textrm{ for } j \leq (n-1) \\
\left(1,1,1\right) & \textrm{ for } j = n \\
\left(j-n+1, j-n, j-n \right) & \textrm{ for } j \geq (n+1)\,, \\
\end{cases}
\end{equation}
and
\begin{equation}
\mathbf{k}_- = 
\begin{cases}
\left(n+j, n+j, n+j-1 \right) & \textrm{ for } j \leq (N-n) \\
\left(N,N,N\right) & \textrm{ for } j = (N+1-n)\\
\left(2N - n -j +1, 2N - n - j +1, 2N-n-j + 2 \right) & \textrm{ for } j \geq (N+2-n)\,. \\
\end{cases}
\end{equation}

Finally, the angular function $J$ is given by:

\begin{align}
\begin{split}
J(x, y) &= x \asin\left(\frac{y-\beta x}{\sqrt{1-x^2}\sqrt{1-\beta^2}}\right)\\
&\, + y \atan\left(\frac{x-y\beta}{\sqrt{t}}\right) \\
&\, + \frac{1}{2} \atan\left(\frac{1 - y^2 - \beta^2 - x + y\beta(1+x)}{(y-\beta) \sqrt{t}}\right) \\
&\, - \frac{1}{2} \atan\left(\frac{1 - y^2 - \beta^2 + x - y\beta(1-x)}{(y+\beta) \sqrt{t}}\right)\,,
\end{split}
\end{align}
where we define $\beta = \vmin/v$ and $t = (1-x^2)(1-\beta^2) - (y - \beta x)^2$. We note that the angular integrals in the Radon transform lead to this analytic form for arbitrary $N$, meaning that only 1-dimensional integrals over $v$ need to be performed. A python implementation of this algorithm is available on request from the author.
\end{appendix}

\bibliographystyle{JHEP}
\bibliography{Directional}

\end{document}